\documentclass[useAMS,usenatbib]{mnras}
\pdfoutput=1 
\usepackage{amsmath,amstext}
\usepackage[T1]{fontenc}
\usepackage{graphicx}
\usepackage{pdflscape}
\usepackage{color}
\usepackage[table,dvipsnames]{xcolor}
\usepackage[hang]{subfigure}
\usepackage{appendix}
\usepackage{array,multirow}
\usepackage{amssymb}
\usepackage{hyperref}
\usepackage{lmodern}
\usepackage{booktabs}
\usepackage{multirow}
\usepackage{comment}
\usepackage{pifont}
\usepackage{tabularx}
\usepackage{makecell}
\usepackage{titlesec}
\usepackage[normalem]{ulem}
\usepackage{xspace}
\usepackage{textgreek}
\usepackage{soul}

\newcommand{\Halpha}{\text{H\textalpha}\xspace}

\definecolor{sunrise}{RGB}{245, 123, 66}

\newcommand{\gonem}{G140M/F070LP\xspace}
\newcommand{\gtwom}{G235M/F170LP\xspace}
\newcommand{\gthreem}{G395M/F290LP\xspace}
\newcommand{\gthreeh}{G395H/F290LP\xspace}

\newcommand{\beagle}{{\footnotesize BEAGLE}\xspace}
\newcommand{\eazy}{{\footnotesize EAZY}\xspace}
\newcommand{\jwst}{\textit{JWST}\xspace}
\newcommand{\hst}{\textit{HST}\xspace}
\newcommand{\targetSelection}{Paper {\footnotesize I}\xspace}

\newcommand{\medhst}{\texttt{mediumhst}\xspace}
\newcommand{\medjwst}{\texttt{mediumjwst}\xspace}
\newcommand{\deephst}{\texttt{deephst}\xspace}
\newcommand{\deepjwst}{\texttt{deepjwst}\xspace}
\newcommand{\ultradeep}{\texttt{ultradeep}\xspace}
\newcommand{\medhstgs}{\texttt{goods-s-mediumhst}\xspace}
\newcommand{\medjwstgs}{\texttt{goods-s-mediumjwst}\xspace}
\newcommand{\medhstgn}{\texttt{goods-n-mediumhst}\xspace}
\newcommand{\medjwstgn}{\texttt{goods-n-mediumjwst}\xspace}
\newcommand{\deephstgs}{\texttt{goods-s-deephst}\xspace}
\newcommand{\deepjwstgs}{\texttt{goods-s-deepjwst}\xspace}
\newcommand{\ultradeepgs}{\texttt{goods-s-ultradeep}\xspace}

\defcitealias{DR1}{DR1}
\defcitealias{DR3}{DR3}
\defcitealias{DR4_paper2}{Paper {\footnotesize II}\xspace}

\title[JADES \textsc{DR4}]{JADES Data Release 4 Paper \textsc{I}: Sample Selection, Observing Strategy and Redshifts of the complete spectroscopic sample}

\author[Curtis-Lake, Cameron, Bunker et al.]{Emma Curtis-Lake,$^{1}$\thanks{E-mail: e.curtis-lake@herts.ac.uk}\thanks{These authors contributed to this work equally.}
Alex J. Cameron,$^{2}$\thanks{E-mail: alex.cameron@physics.ox.ac.uk}$\dagger$  
Andrew J.\ Bunker,$^{2}$\thanks{E-mail: andy.bunker@physics.ox.ac.uk}$\dagger$ 
Jan Scholtz,$^{3,4}$
Stefano Carniani,$^{5}$
\newauthor
Eleonora Parlanti,$^{5}$
Francesco D'Eugenio,$^{3,4}$
Peter Jakobsen,$^{6,7}$
Christopher N. A. Willmer,$^{8}$
\newauthor
Santiago Arribas,$^{9}$
William M. Baker,$^{10}$
Stéphane Charlot,$^{11}$
Jacopo Chevallard,$^{2}$
\newauthor
Chiara Circosta,$^{12}$
Mirko Curti,$^{13}$
Qiao Duan,$^{3,4}$
Daniel J.\ Eisenstein,$^{14}$
Kevin Hainline,$^{8}$
Zhiyuan Ji,$^{8}$
\newauthor
Benjamin D.\ Johnson,$^{14}$
Gareth C. Jones,$^{3,4}$
Roberto Maiolino,$^{3, 4, 15}$
Michael V. Maseda,$^{16}$
\newauthor
Michele Perna,$^{9}$
Pablo G. Pérez-González,$^{9}$
Tim Rawle,$^{17}$
Marcia Rieke,$^{8}$
Pierluigi Rinaldi,$^{18}$
\newauthor
Brant Robertson,$^{19}$
Bruno Rodr\'iguez Del Pino,$^{9}$
Aayush Saxena,$^{2, 15}$
Irene Shivaei,$^{9}$
Renske Smit,$^{}$
\newauthor
Sandro Tacchella,$^{3,4}$
Hannah \"Ubler,$^{21}$
Giacomo Venturi,$^{5}$
Christina C. Williams,$^{22}$
and Chris Willott$^{23}$
\\
$^{1}$Centre for Astrophysics Research, Department of Physics, Astronomy and Mathematics, University of Hertfordshire, \\ ~ ~ Hatfield AL10 9AB, UK\\
$^{2}$Department of Physics, University of Oxford, Denys Wilkinson Building, Keble Road, Oxford OX1 3RH, UK\\
$^{3}$Kavli Institute for Cosmology, University of Cambridge, Madingley Road, Cambridge, CB3 OHA, UK\\
$^{4}$Cavendish Laboratory - Astrophysics Group, University of Cambridge, 19 JJ Thomson Avenue, Cambridge, CB3 OHE, UK\\
$^{5}$Scuola Normale Superiore, Piazza dei Cavalieri 7, I-56126 Pisa, Italy\\
$^{6}$Cosmic Dawn Center (DAWN), Copenhagen, Denmark\\
$^{7}$Niels Bohr Institute, University of Copenhagen, Jagtvej 128, DK-2200, Copenhagen, Denmark\\
$^{8}$Steward Observatory, University of Arizona, 933 N. Cherry Avenue, Tucson, AZ 85721, USA\\
$^{9}$Centro de Astrobiolog\'ia (CAB), CSIC–INTA, Cra. de Ajalvir Km.~4, 28850- Torrej\'on de Ardoz, Madrid, Spain\\
$^{10}$DARK, Niels Bohr Institute, University of Copenhagen, Jagtvej 155A, DK-2200 Copenhagen, Denmark\\
$^{11}$Sorbonne Universit\'e, CNRS, UMR 7095, Institut d'Astrophysique de Paris, 98 bis bd Arago, 75014 Paris, France\\
$^{12}$Institut de Radioastronomie Millimétrique (IRAM), 300 Rue de la Piscine, 38400 Saint-Martin-d'Hères, France\\
$^{13}$European Southern Observatory, Karl-Schwarzschild-Strasse 2, 85748 Garching, Germany\\
$^{14}$Center for Astrophysics $|$ Harvard \& Smithsonian, 60 Garden St., Cambridge MA 02138 USA\\
$^{15}$Department of Physics and Astronomy, University College London, Gower Street, London WC1E 6BT, UK\\
$^{16}$Department of Astronomy, University of Wisconsin-Madison, 475 N. Charter St., Madison, WI 53706 USA\\
$^{17}$European Space Agency (ESA), European Space Astronomy Centre (ESAC), Camino Bajo del Castillo s/n, 28692 Villafranca del \\ ~ ~  Castillo, Madrid, Spain\\
$^{18}$Space Telescope Science Institute, 3700 San Martin Drive, Baltimore, Maryland 21218, USA\\
$^{19}$Department of Astronomy and Astrophysics University of California, Santa Cruz, 1156 High Street, Santa Cruz CA 96054, USA\\
$^{20}$Astrophysics Research Institute, Liverpool John Moores University, 146 Brownlow Hill, Liverpool L3 5RF, UK\\
$^{21}$Max-Planck-Institut f\"ur extraterrestrische Physik (MPE), Gie{\ss}enbachstra{\ss}e 1, 85748 Garching, Germany\\
$^{22}$NSF National Optical-Infrared Astronomy Research Laboratory, 950 North Cherry Avenue, Tucson, AZ 85719, USA\\
$^{23}$NRC Herzberg, 5071 West Saanich Rd, Victoria, BC V9E 2E7, Canada\\
\vspace{-1.5 cm}
}

\begin{document}

\date{}

\pagerange{\pageref{firstpage}--\pageref{lastpage}} \pubyear{2025}

\maketitle

\label{firstpage}
\begin{abstract}
\noindent 
This paper accompanies Data Release 4 of the \jwst Deep Extragalactic Survey (JADES), which presents the full NIRSpec spectroscopy of the survey. We provide spectra of 5190 targets across GOODS-North and GOODS-South (including the \textit{Hubble} Ultra Deep Field), observed with the low-dispersion ($R \approx 30$--$300$) prism and three medium-resolution ($R \approx 1000$) gratings spanning $0.8 < \lambda < 5.5~\mu$m; 2654 were also observed with the higher-resolution ($R \approx 2700$) G395H grating. The tiered survey design obtained $\gtrsim 20$~hr exposures for $\sim700$ galaxies in the Deep and Ultra Deep tiers, and shallower observations ($\sim1$--3~hr per setting) of $>4400$ galaxies in the Medium tiers. Targets were selected from photometric redshifts or colours, with priority given to rest-UV-selected galaxies at $z > 5.7$ and F444W-selected galaxies at $1.5 < z < 5.7$. We describe the full target selection and present spectroscopic redshifts and success rates. In total we obtain robust redshifts for 3297 galaxies, including 396 at $z > 5.7$ and 2545 at $1.5 < z < 5.7$. To facilitate uniform analyses, we define `gold' sub-samples based on UV- and F444W-selection. Using the parent samples and redshift success rates, we construct rest-UV luminosity functions at $6 \lesssim z \lesssim 9$ from the Medium- and Deep-\jwst tiers. Our number densities agree well with previous determinations from both photometric and spectroscopic samples, with modest interloper fractions confirming the reliability of photometric UV-bright galaxy selections at these redshifts.
\end{abstract}

\begin{keywords}
galaxies: evolution, galaxies: high-redshift, galaxies: general, catalogues, surveys, techniques: spectroscopic, methods: observational
\vspace{-1.5 cm}
\end{keywords}

\fixfootnotes

\section{Introduction}

The journey from initial bold concept to construction, launch and new scientific discoveries with \jwst has spanned three decades.  Now, the telescope is performing exceptionally \citep{Rigby2023,Rieke2023_NIRCam}. In particular, the myriad spectroscopic capabilities have opened up brand new discovery space for the study of early galaxy evolution. Thanks to the incredibly sensitive Near-InfraRed Spectrograph \citep[NIRSpec;][]{Jakobsen2022}, candidate galaxies inhabiting the Universe within the first few billion years are being confirmed in record numbers, and their diverse 
properties are being revealed.  Thanks to deep spectroscopy, black holes have been found in the early Universe \citep[e.g.][]{Kokorev2023, Maiolino2024}, intriguing chemical abundances have been identified that are challenging to explain \citep[e.g.][]{Cameron2023_Nitrogen, DEugenio2024_Carbon, Curti2024_GS_z9, Schaerer2024, Topping2025}, candidates identified by NIRCam that are more distant than  could be identified with \textit{Hubble Space Telescope} (\hst) have been confirmed \citep[e.g.][]{Curtis-Lake2023, Wang2023, Castellano2024, Carniani2024_GS_z14, Witstok2025_z13, Naidu2025},
and indications of very diverse star-formation histories have been uncovered \citep[e.g.][]{Boyett2024, Endsley2024, Looser2024, Covelo-Paz2025}.

Two of the four originally scoped science goals for the \jwst mission were `first light and reionisation' and `the assembly of galaxies' \citep{Gardner2006_JWST}, both of which are encapsulated in the motivations for the \jwst Advanced Deep Extragalactic Survey \citep[JADES;][]{Eisenstein2023_JADES}.  The survey was designed to provide legacy imaging and spectroscopy in the GOODS (Great Observatories Origins Deep Survey; \citealt{GOODS_2004}) North and South fields. These fields already offer extensive multi-wavelength coverage and some of the deepest \hst\ imaging \citep{Illingworth2013,Koekemoer2013_UDF}, including the \textit{Hubble} Ultra Deep Field \citep[HUDF;][]{Beckwith2006_HUDF}.  Developed through a collaboration between the NIRCam (Near-Infrared Camera; \citealt{Rieke2023_NIRCam}) and NIRSpec instrument science teams, the survey consists of NIRCam imaging designed to enable follow-up of \jwst-detected sources with NIRSpec, while also making efficient use of parallel observations.
JADES represents a significant fraction of the Guaranteed Time Observations (GTO) of the NIRCam and NIRSpec teams, and is the largest \jwst programme in the first two cycles. The NIRSpec component of JADES is complemented by other NIRSpec GTO programmes, such as the Wide spectroscopic tier \cite{Maseda2024} which covers a greater area over more fields but to shallower depth, and the Galaxy Assembly NIRSpec Integral Field Spectroscopy survey, \citep[GA-NIFS, e.g.][]{Ubler2024,DEugenio2024,Parlanti2025,Perna2025} which obtains detailed spatially-resolved spectra over a $3"\times3"$ field.  Also notable is the NIRSpec component of the Systematic Mid-Infrared Instrument (MIRI) Legacy Extragalactic Survey \citep[SMILES;][]{Rieke2024,Alberts2024,Zhu2025}, mainly targeting cosmic noon galaxies and obscured AGNs in the HUDF, with multi-band MIRI imaging.

The aim for the spectroscopic side of the JADES survey was to compile a statistically representative sample of galaxy spectra from cosmic noon 
to well within the epoch of reionisation, while also targetting the most distant galaxy candidates at cosmic dawn \citep{Bunker2020}. NIRSpec is unique in its spectroscopic capabilities in that it has a multi-object mode that uses a micro-shutter array \citep{Ferruit2022}. Although this mode makes efficient use of a high-degree of multiplexing, observing $\gtrsim$150 targets simultaneously, decisions must be made about which galaxies falling within the field of view will be targetted for spectroscopy. The survey selection function is complicated but driven by the wish to understand the star formation history of the Universe, and the assembly of stellar mass.

The JADES survey is now complete and the full spectroscopic sample is presented to the community in this paper and \cite{DR4_paper2} \citepalias[hereafter][]{DR4_paper2}, which collectively form Data Release 4 (DR4)\footnote{Available on the JADES website \url{https://jades-survey.github.io/scientists/data.html} or \url{https://archive.stsci.edu/hlsp/jades}}.  
In this paper, (\targetSelection), we present the target selection strategies, 
measured redshifts and selection success rates. In addition, due to the complicated nature of the selection function, we define two subsets for statistical galaxy-evolution studies: one selected to an apparent magnitude limit in the reddest filter, and the other comprising high redshift galaxies selected through their Lyman break, down to an apparent magnitude limit sampling the rest-frame ultra-violet (UV).
\citetalias{DR4_paper2} will present the data reduction using the guaranteed-time-observations (GTO) team pipeline 
and the emission line measurements and properties.
The final imaging products will be made available in a forthcoming Data Release 5 (Johnson et al., in prep., Robertson et al., in prep.). 

In Section~\ref{sec:overview}, we provide a brief overview of the JADES spectroscopic survey before describing what is new in this release and the design of the observing program.
Section~\ref{sec:target_selection} then outlines how targets were selected, including the assembly of photometric catalogues, the criteria used to divide targets into priority classes, and how MSA slitlets were allocated to targets. 
In Section~\ref{sec:redshift_success}, we present the success rates of our target selection based on spectroscopic redshift measurements obtained from the survey.  Section~\ref{sec:gold_sample} presents our two ``gold'' sub-samples, designed to facilitate statistical galaxy evolution studies. Finally, a brief summary is given in Section~\ref{sec:summary}. Throughout this paper we assume a flat $\Lambda$CDM cosmology with $H_0 = 67.66$ km s$^{-1}$ Mpc$^{-1}$ and $\Omega_M = 0.310$ and adopt the AB magnitude system \citep{OkeGunn1983}.

\section{The NIRSpec component of the JADES survey}
\label{sec:overview}

\begin{figure*}
    \centering
    \includegraphics[width=0.32\linewidth]{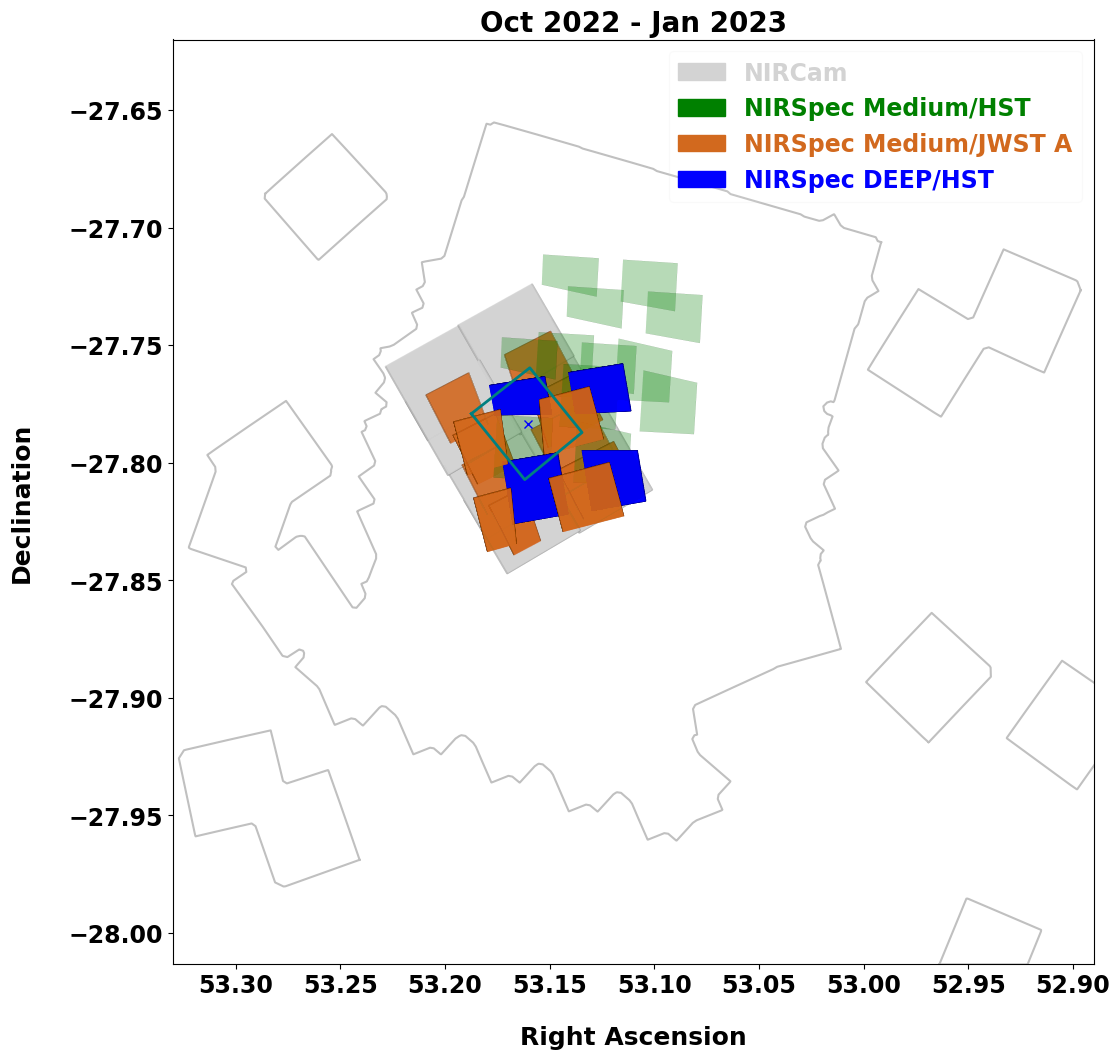}
    \includegraphics[width=0.32\linewidth]{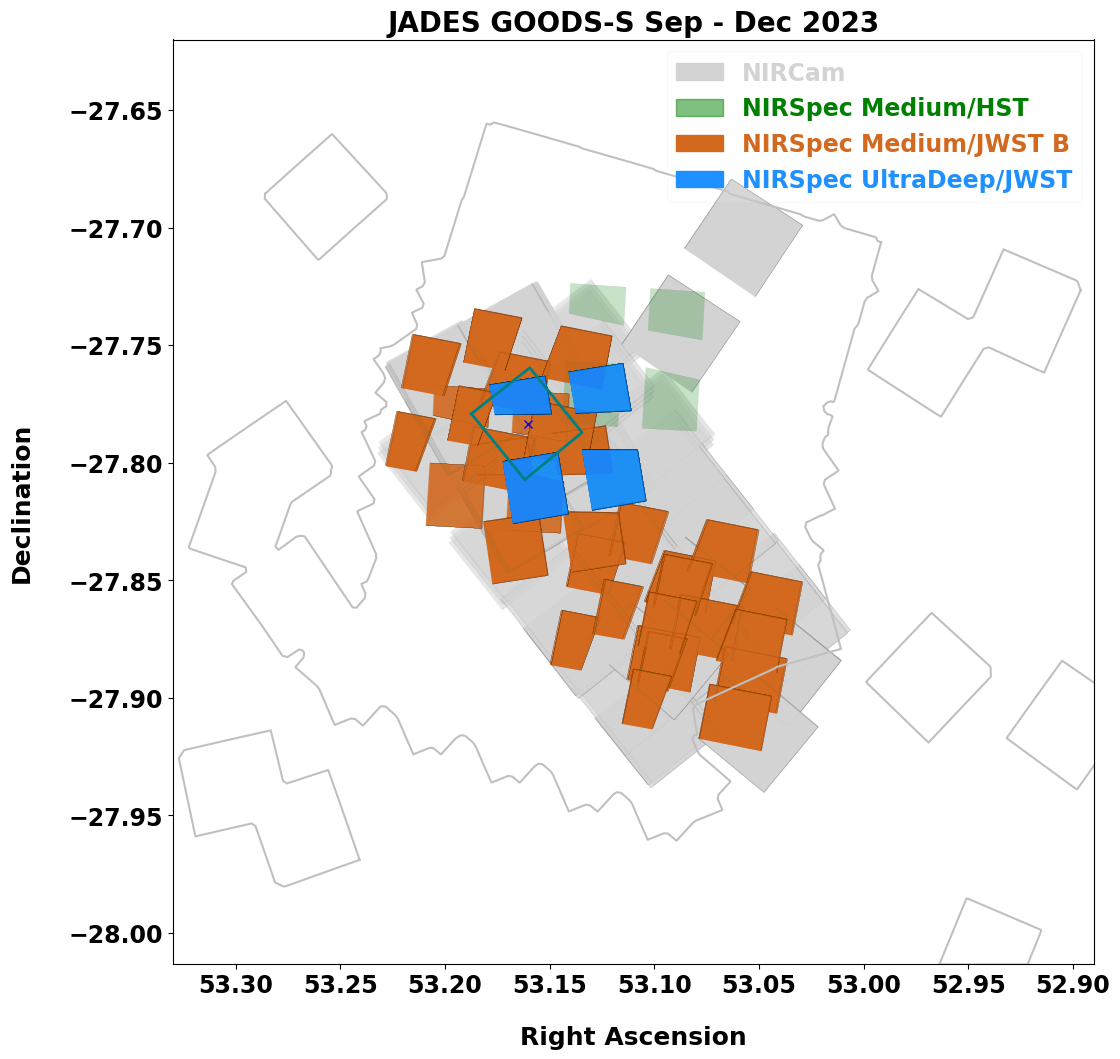}
    \includegraphics[width=0.32\linewidth]{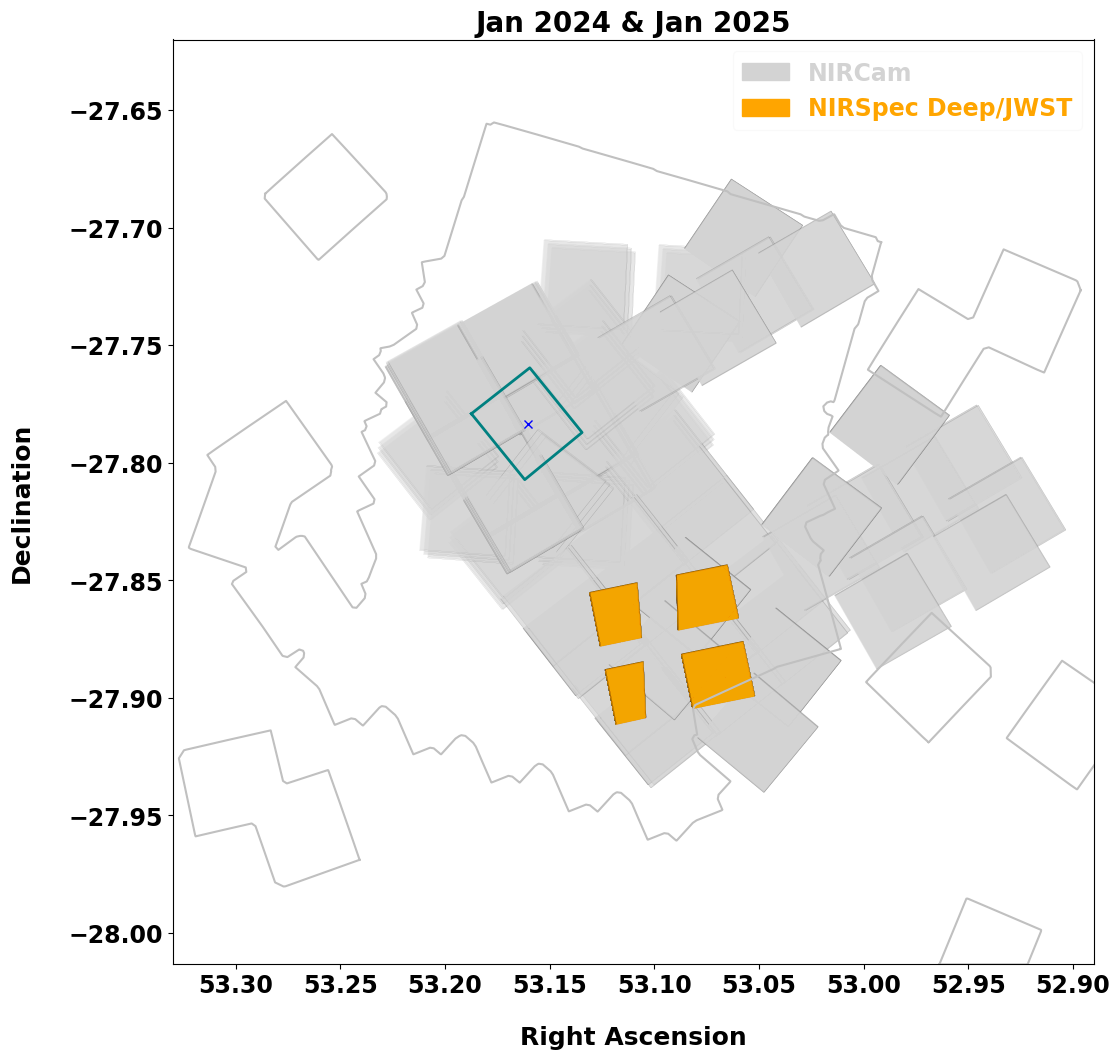}
    \caption{Timeline of the JADES survey footprint in GOODS-South. The grey outline shows the coverage of pre-existing HST data in the WFC3 F160W filter taken from the Hubble Legacy Fields \citet{Whitaker2019}.  The Green rectangle and plus sign indicate the outline and central position of the \textit{Hubble} eXtreme Deep Field \citep{Illingworth2013}. The filled grey regions show the JADES NIRCam data available for NIRSpec target selection undertaken within the time-frame indicated at the top of each panel. NIRSpec pointings taken within the same time-frame are shown with the shading set by the Tier, as indicated in the legend.  The regions for each pointing cover the useable NIRSpec MSA area that avoids truncated PRISM spectra, and indicate the area of the MSA used for target allocation. The Medium/\jwst pointings in the Oct 2022 - Jan 2023 (left) panel indicate the early \medjwstgs pointings that constitute the subsample labelled `GSa' that were included in \citetalias{DR3}, while the Medium/\jwst pointings in the Sep - Dec 2023 (central) panel constitute the \medjwstgs pointings referred in the text as `GSb' and are included in this current data release (see Section~\ref{sec:target_selection} for more details). The blue shaded region in the left panel shows the Deep/HST NIRSpec pointing that comprised the \citetalias{DR1} release. Other than the UltraDeep/JWST pointing in the central panel, all pointings in the central and right panels are included in this release for the first time.}
    \label{fig:footprint_gs}
\end{figure*}

\begin{figure*}
    \centering
    \includegraphics[width=0.32\linewidth]{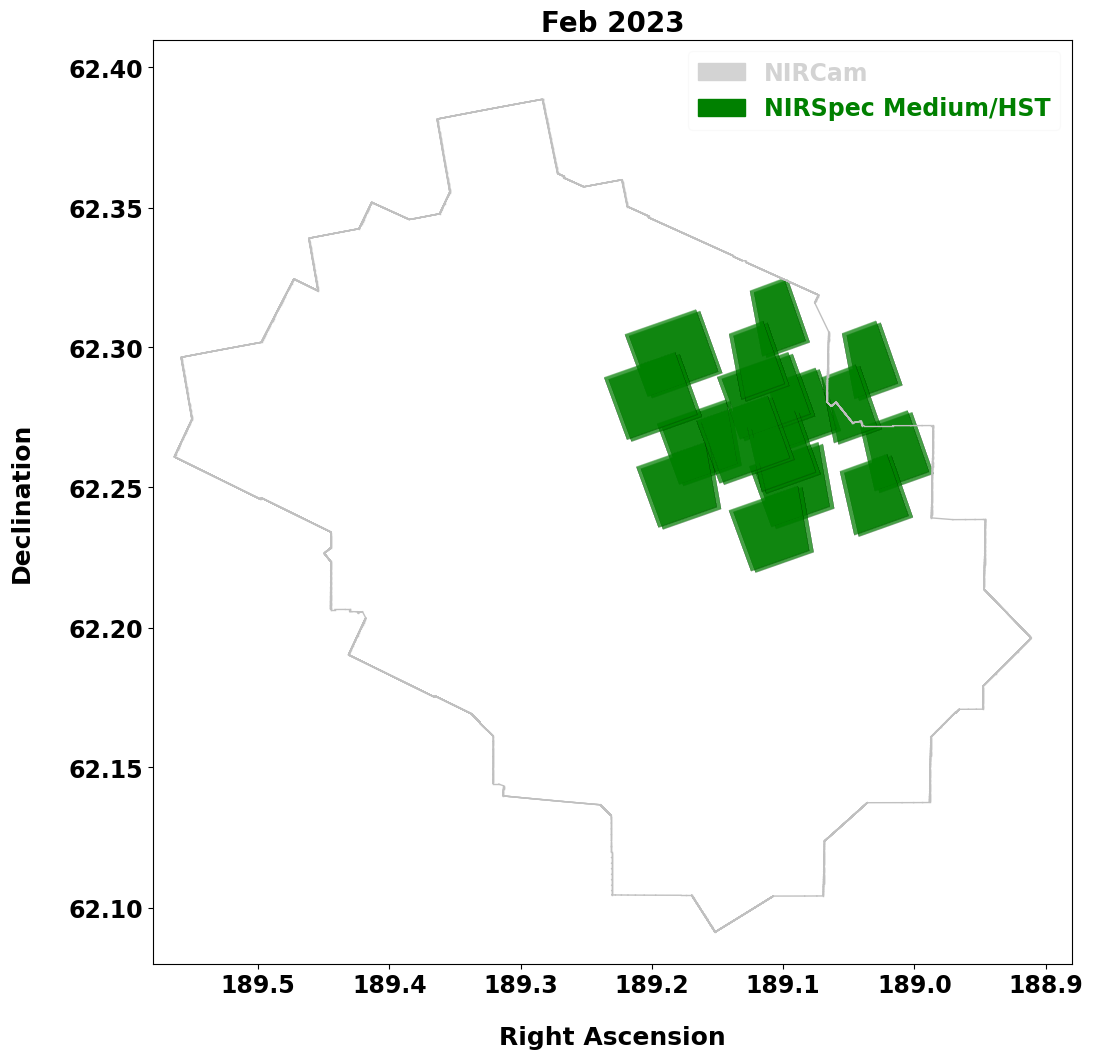}
    \includegraphics[width=0.32\linewidth]{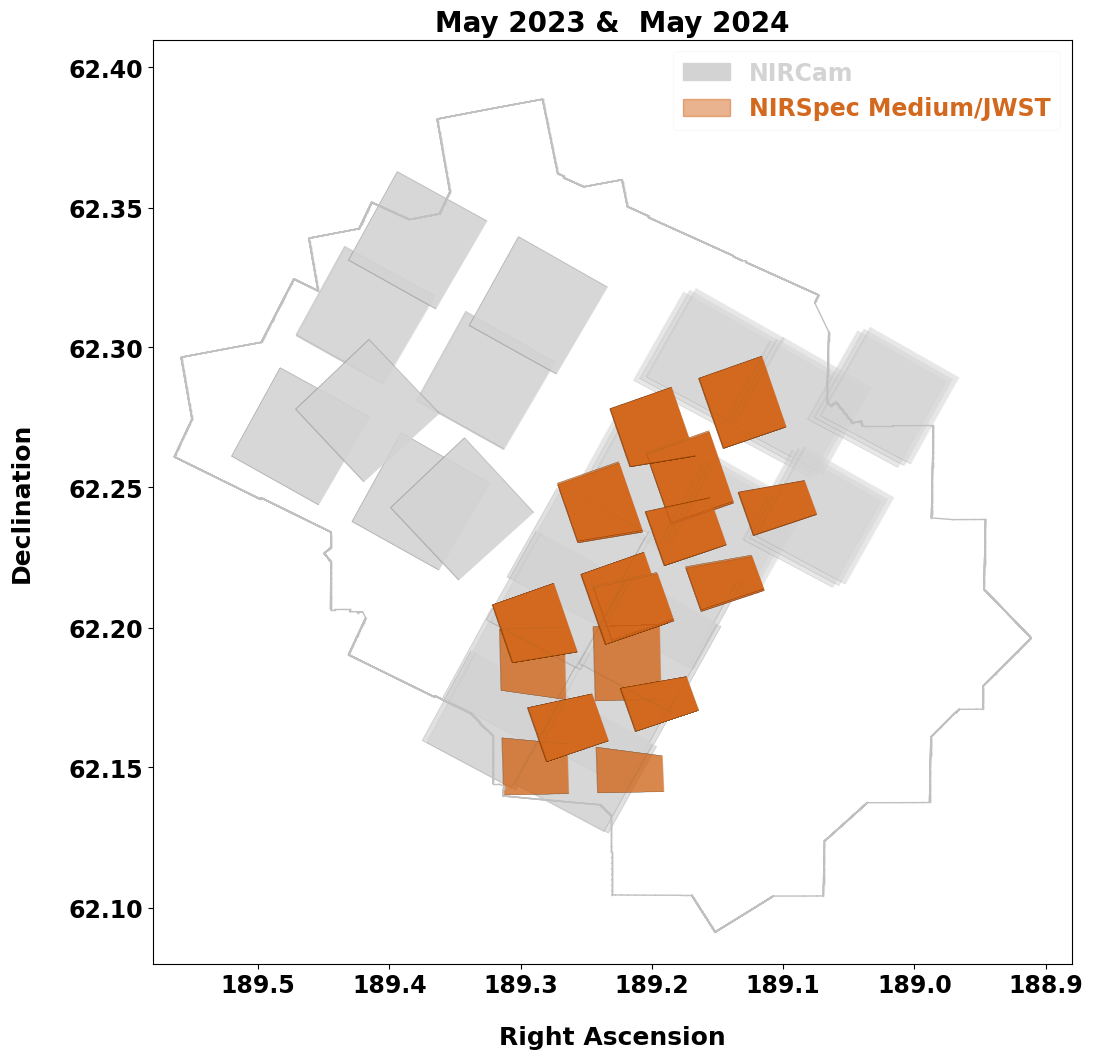}
    \caption{As for Figure~\ref{fig:footprint_gs}, but showing the timeline of the JADES survey footprint in GOODS-North.  As in the previous figure, the grey outline represents the footprint of prior \hst WFC3 F160W observations combined by \citet{Illingworth2017}. These pointings were included in the \citetalias{DR3} release.}
    \label{fig:footprint_gn}
\end{figure*}

\subsection{Overview}

The JADES survey consists of NIRCam imaging and NIRSpec spectroscopy in the GOODS-S and GOODS-N fields \citep[PIDs 1180, 1181, 1210, 1286, 1287, 3215; PIs Eisenstein, L\"utzgendorf;][]{Eisenstein2023_JADES, Eisenstein2023_JOF}.  The spectroscopic component of the JADES survey consisted of a tiered 
strategy which, broadly speaking, obtained very deep spectroscopy on a few hundred targets and medium-depth spectroscopy of several thousand targets, balancing depth and sample size within the available observing time.

In each of the JADES observations, simultaneous imaging and spectroscopy were carried out with NIRCam and NIRSpec in coordinated parallel mode\footnote{Note that some observations in PIDs 1180 \& 1181 consisted of NIRCam imaging with MIRI imaging in parallel, but this paper refers to the spectroscopic data release which was comprised of observations where NIRCam and NIRSpec were operating simultaneously.}, with the two instruments separated by a few arcmin in the focal plane. 
While in terms of \jwst operations, NIRSpec is always the prime instrument, the outline for part of the survey was driven by desire to construct contiguous imaging mosaics with NIRCam. As a result, for these observations, NIRCam drove the location and orientation of the pointings, which we refer to hereafter as ``NIRCam prime''. For the remaining observations, NIRSpec was given more freedom to follow-up the most promising candidates; we refer to these as ``NIRSpec prime''. 
Figures~\ref{fig:footprint_gs} and~\ref{fig:footprint_gn} show the development over time of the footprint of the NIRCam mosaic, and the corresponding NIRSpec coverage in GOODS-S and GOODS-N, respectively. 

The original JADES spectroscopic programme, which used GTO time apportioned to the NIRSpec-IST (Instrument Science Team), envisaged two `deep' pointings in GOODS-S. For these deep pointings, NIRSpec was prime while NIRCam was used in parallel. 
The first deep pointing was planned for early in the lifetime of \jwst to test the capabilities of NIRSpec at long exposure times.  The targets were primarily selected from \hst imaging, though fortunate timing\footnote{A late MSA re-design was required following issues with MSA shorts.} allowed the addition of some NIRCam-selected targets \citep{Curtis-Lake2023, Robertson2023}. Spectroscopy from this \deephstgs pointing was released before the end of the proprietary time and presented in the first data release paper, DR1 \citep{DR1} \citepalias[hereafter][]{DR1}.  The second deep spectroscopic pointing was planned once deep NIRCam imaging was available to enable follow-up of newly identified targets \citep[e.g.][]{Carniani2024_GS_z14}. This second \deepjwstgs pointing (PID 1287) is presented in full for the first time in this paper.

In addition to the JADES GTO survey, 
an \ultradeep spectroscopic pointing was observed in GOODS-S through a GO programme \citep[PID 3215,][]{Eisenstein2023_JOF}, which formed part of the third JADES data release \citep[][hereafter \citetalias{DR3}]{DR3}. The \ultradeep MSA footprint repeated that of the \deephst .

Our medium-depth observations were a combination of NIRCam prime and NIRSpec prime. In each field (GOODS-S and GOODS-N), our initial observations focussed on constructing a contiguous NIRCam mosaic.
This had two consequences: (1) the position and orientation of our NIRSpec pointings had very little freedom, and (2) our spectroscopic targets for these observations had no available NIRCam pre-imaging. 
Thus, target selection for this \medhst tier was based on \hst photometry.
In GOODS-S, this comprised 12 pointings of \medhstgs (PID: 1180) and in GOODS-N a further 8 pointings of \medhstgn (PID: 1181).

Our subsequent visits, with NIRSpec as prime, went deeper in exposure time and consisted of an originally-planned 12 pointings (8 in GOODS-S, PID 1286, and 4 in GOODS-N, PID 1181), designed to observe targets selected based on the NIRCam imaging (\medjwst).

The \medhstgs observations 
were heavily affected by electrical short circuits in the MSA  producing bright regions in the detector \citep{Bechtold2024}, and around two-thirds the spectra were severely contaminated \citepalias[see Appendix~A in ][]{DR3}. 
The re-observations of these were planned during a time of the year when the spacecraft had a different roll angle which required the MSA configurations to be re-designed. Furthermore, since NIRCam imaging was now available, we decided it was better to use this time to expand the \medjwst tier.  Instead of repeating the initial setup, three additional \medjwstgs pointings were designed with a slightly modified exposure strategy (see Section~\ref{subsec:observations}) but using the same target selection approach as the main \medjwst pointings.

 \subsection{What's new in DR4}
 \label{subsec:whats_new}

In \citetalias{DR1}, we released the \deephstgs\ spectroscopy alongside our first area of NIRCam imaging \citep{Rieke2023_DR1}, while DR2 comprised only NIRCam imaging. 
In \citetalias{DR3} we released all spectra observed before September 2023, plus the \ultradeepgs\ spectra observed in October 2023. This included a large number of short-affected exposures in \medhstgs\ which, despite being usable for determining redshifts, did not have reliable flux calibration, due to heavy contamination from the bright glow from electrical short circuits in the MSA array \citep{Rawle2022}.

In DR4, we release all the remaining JADES spectroscopic data and re-reductions of spectra previously released in \citetalias{DR1} \& \citetalias{DR3} which have been processed with the latest updates to the pipeline and new calibration files \citepalias[see][]{DR4_paper2}. 
Since the purpose of this data release is to provide science-ready data products, we have included only the 677 uncontaminated spectra from the heavily short-affected \medhstgs field. The other 665 spectra, although released in DR3, were excluded because their flux calibration is not considered reliable.

In addition to re-reductions of previously released data, we also present several new NIRSpec pointings in DR4 that have not appeared in previous JADES data releases.
These include 7 new pointings (1321 Prism spectra) in the \medjwstgs tier and the second deep pointing, \deepjwstgs (235 Prism spectra). 
We note that one of the three dithers in \deepjwstgs failed during the initial observation window in January 2024, and was re-taken in January 2025. This data release includes the full planned exposure time for this tier. 
Additionally, one of the original pointings in \medjwstgn was skipped due to a guiding error, and the first set of re-observations was affected by MSA shorts.  The final data were re-taken in May 2024. These are now included in this release, contributing 237 new unique Prism spectra, as well as additional depth to some previously released spectra.

In summary, this release includes (i) new spectra from 1794 targets, and (ii) re-reductions of previously released spectra from 3425 targets, while omitting 665 short-affected spectra that had been released in \citetalias{DR3}.


\begin{table*}
\scriptsize
\begin{center}
\caption{
Summary of the observational setup and number of targets in each tier of the survey.  
}
\label{tab:obssummary}

\begin{tabular}{llcccccccccccc}
\hline
PID & Field / Tier name & Selection & Obs Date & PRISM depth$^{\rm a}$ & Grating depth$^{\rm a}$ & G395H? & \multicolumn{3}{c}{Number of targets$^{\rm c}$} \\
 & &  &  & h & h & & Pr & Pr \& Gr & Gr-only  \\
\hline
\multicolumn{10}{c}{ }\\
\multicolumn{10}{l}{\bf ~~~~~~~~~~GOODS-South}\\
\hline
1210 & \verb|goods-s-deephst| & \hst/\jwst & Oct-22 & 9.2 -- 27.7 & 2.3 -- 6.9  & \checkmark & 253 & 198 &  \\
1287 & \verb|goods-s-deepjwst| & \jwst & Jan-24 \& Jan-25 & 9.2 -- 27.7 & 2.3 -- 6.9  & \checkmark  & 235 & 215 & \\
3215 & \verb|goods-s-ultradeep| & \jwst & Oct-23 & 16.2 -- 50.8 & 18.5 -- 37.0$^{\rm b}$ & & 228 & 189 & \\
\hline
1180 & \verb|goods-s-mediumhst| & \hst & Oct-22 & 1.0--4.1 & 0.9--3.4 & & 677 & 600 & 1 \\
     & \verb|goods-s-mediumjwst| & \jwst & Jan-23 \& Oct-23 & 1.4--5.2 & 1.7--4.3 & & 533 & 513 & 1 \\
1286 & \verb|goods-s-mediumjwst| & \jwst & Jan-23 & 1.5--9.2 & 1.6--9.5 & \checkmark & 169 & 159 &  \\
     &                           &      & Oct-23 & 1.5--9.2 & 1.6--9.5 & \checkmark & 206 & 193 & 3 \\
     &                           &      & Dec-23 & 1.5--9.2 & 1.6--9.5 & \checkmark & 1115 & 997 & 1 \\
\multicolumn{10}{c}{ }\\
\multicolumn{10}{l}{\bf ~~~~~~~~~~GOODS-North}\\
\hline
1181 &  \verb|goods-n-mediumhst| &   \hst & Feb-23 & 1.7--6.8 & 0.9--3.4 & & 853 & 711 & 76 \\
 & \verb|goods-n-mediumjwst| &  \jwst & May-23 \& May-24 & 0.9--7.7 & 0.9--7.7 & \checkmark & 950 & 892 & 3 \\
\hline
\end{tabular}
\end{center}
\raggedright
$^{\rm a}$ Exposure times given are the 25th percentile and maximum value among all targets that received non-zero exposure time in the given tier. These are designed simply to give a sense for the depths.\\
$^{\rm b}$ Unlike other tiers, which had equal exposure time across all medium resolution gratings, \texttt{goods-s-ultradeep} received very deep exposures in G395M/F290LP (values shown), somewhat shallower exposures in G140M/F070LP (values shown divided by four), and no exposure in G235M/F170LP.\\
$^{\rm c}$ Pr: Has Prism spectrum; Pr \& Gr: Prism and grating; Gr: grating only (no Prism due to failed observation).
\end{table*}


\subsection{Observations}
\label{subsec:observations}

\begin{figure}
    \centering
    \includegraphics[width=0.99\linewidth]{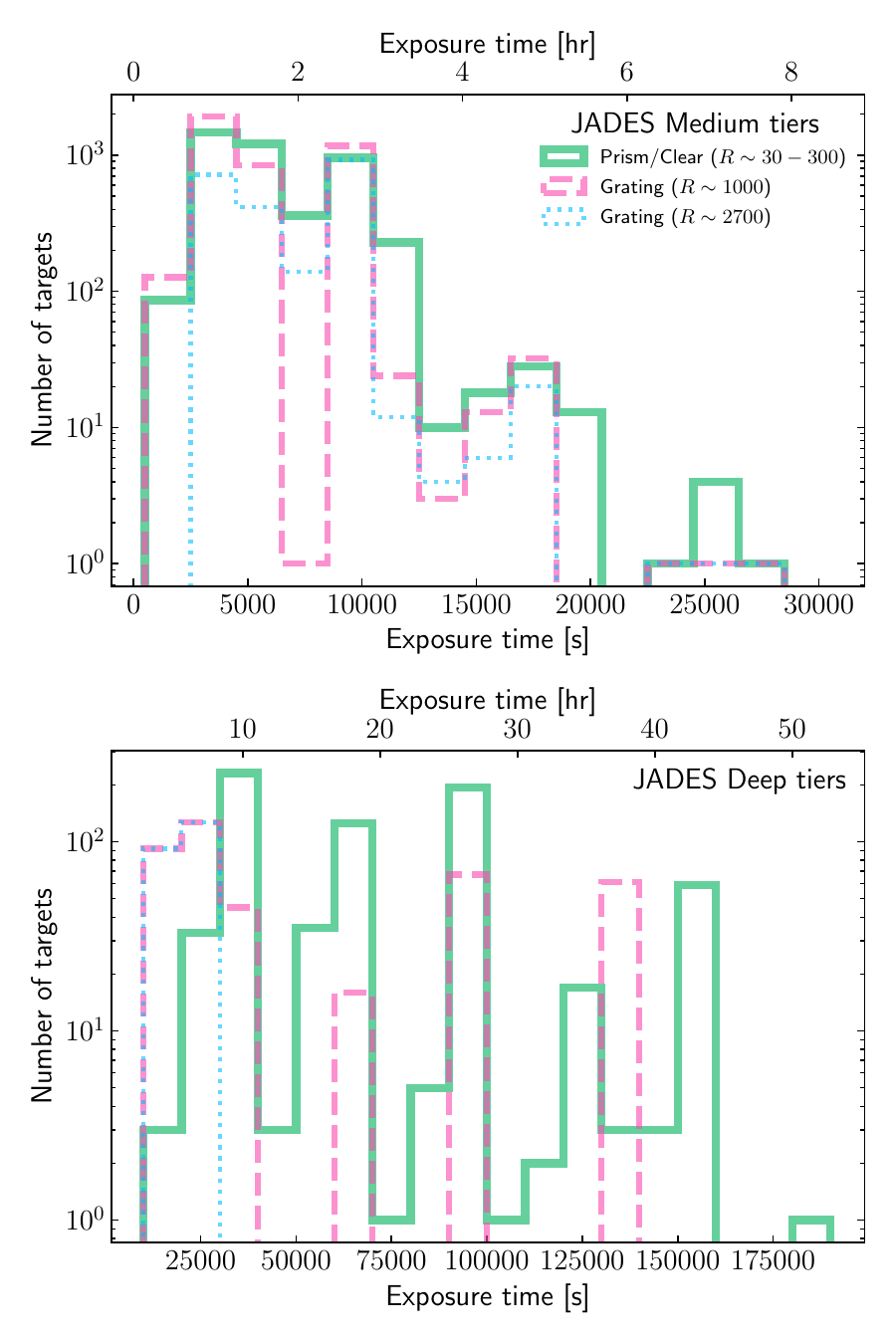}
    \caption{Histogram of exposure times from the full JADES survey. The top panel shows exposure times from the ``Medium'' tiers (see text for description). The green solid line shows exposures in the Prism/Clear mode. Pink dashed line shows the exposure time in each medium resolution ($R\sim1000$) gratings (note that these targets received the given exposure time in each of G140M/F070LP, G235M/F170LP, G395M/F290LP to cover the full wavelength range). The cyan dotted line shows the exposure time in the high resolution ($R\sim2700$) grating G395H/F290LP. The lower panel shows the same, but for the JADES ``Deep'' tiers where for \ultradeepgs, the $R\sim1000$ grating observations are only taken with G395M/F290LP.
    }
    \label{fig:exposure_times}
\end{figure}

The tiers, exposure times and grating/filter combinations for the NIRSpec observations are summarised in Table~\ref{tab:obssummary}.  All observations employed a three-nod pattern for background subtraction, where we move the targets by $\pm$1\,microshutter ($0\farcs 53$) perpendicular to the dispersion direction.

Our two deep tiers, \texttt{goods-s-deephst} (PID: 1210) and \texttt{goods-s-deepjwst} (PID: 1287) 
shared identical observational design which employed three dither positions (`dithers' are spatial offsets to place spectra on different areas of the detectors).  At each dither, 9.3 h of exposure were obtained with Prism/Clear, providing high S/N continuum plus emission line spectra for many sources, albeit at low spectral resolution ($R\approx30-300$). The same three dither positions were then used to obtain full wavelength coverage at $\mathrm{R}\sim1000$, with 2.3 h of exposure per dither in each of \gonem, \gtwom and \gthreem, to resolve closely-spaced emission lines.  Finally, a higher spectral resolution \gthreeh\ 2.3 h  exposure at each dither enables kinematic studies and searches for broad emission line features from unobscured AGN in a subset of sources. 
The three dither positions were selected to maximise target commonality, so as to prioritize increased depth rather than observing more targets, and were $\lesssim1$ arcsecond apart from each other. Therefore, many targets received up to 27 h in Prism and 7 h in each grating. 
\deephstgs was observed in October 2022 and released in \citetalias{DR1}. Two of the three \deepjwstgs dithers were observed in January 2024. The third was skipped because of a guide star acquisition failure and was observed a year later in January 2025.

This release also includes data from the Cycle 2 GO program 3215 as \ultradeepgs \citep{Eisenstein2023_JOF}. This was designed to focus on depth in Prism/Clear and \gthreem. This comprised four dither positions\footnote{It was originally designed to be 5 positions, but as detailed in \citetalias{DR3}, one visit suffered from a bright MSA short.  The re-scheduled observations were re-designed following a new, high-multiplex strategy and presented in  \cite{DEugenio2025}.} of 9.3 h in Prism/Clear and \gthreem, and 2.3 h in \gonem, with no exposure in either of \gtwom\ or \gthreeh.
This covered the same pointing as \deephstgs, but was observed one year later (October 2023) with the same 
top priority targets (class 1, see Section~\ref{subsec:classes_1_6}), but otherwise a different prioritization strategy for the remaining targets.

For our \medhst\ pointings, which had NIRCam in prime, the NIRSpec observations comprised 12 (8) pointings in GOODS-South (GOODS-North). Each 
pointing was 
observed for 2.09 h with the Prism (1.73 h in GOODS-N) and 1.73 h with the \gonem, \gtwom, and \gthreem, with no high-resolution observations. Most targets were observed in either 1 or 2 of these pointings, but a small number fell into overlapping regions and could be observed in multiple pointing pairs, meaning exposure times of up to 6.8 h were obtained with the Prism. 
As mentioned in Section~\ref{subsec:whats_new}, 8 of the 12 pointings in \medhstgs were affected by shorts and are not included in this release.
One pointing in \medhstgn had its Prism observations lost to shorts, meaning there are 76 targets in this tier with grating observations, but no prism observations.

For the \medjwst pointings with NIRSpec as prime, 
the observations employed three dithers in each of Prism/Clear, \gonem, \gtwom, \gthreem\ and \gthreeh with 0.8 h exposures per dither in GOODS-S, and marginally longer 0.86 h exposures per dither in GOODS-N. 
GOODS-N was mostly observed in May 2023. Obs 8 was scheduled for early May 2023, but was skipped due to a  failed guide star acquisition. It was rescheduled for late May 2023 when it was partially completed, but some exposures failed due to shorts, and were observed in May 2024.
GOODS-S was observed in three main epochs. Two out of eight pointings were brought forward to January 2023. One of these failed and was rescheduled and redesigned for October 2023. The remaining six pointings were all observed in December 2023.

Three supplementary \medjwstgs pointings that were planned following the loss of \medhstgs data to shorts had a slightly different and non-uniform exposure setup.  Two pointings consisted of 2 dithers, while the third comprised 3 dithers.  Each dither was observed in Prism/Clear (1 h), \gonem, \gtwom, \gthreem (0.86 h each).  They were originally 
planned for January 2023, but two pointings were heavily affected by MSA shorts and were re-observed in October 2023.

\subsubsection{Pointing selection}

All JADES programmes employed the eMPT software suite \citep{bonaventura_empt_2023} to design the MSA masks, with the observations subsequently imported into the APT. 
In programmes in which NIRSpec was prime the eMPT's Initial Pointing Algorithm (IPA) module was used to first find the set of pointings that provided the largest number of Priority Class 1 targets in the input catalog (described in Section~\ref{sec:target_selection}) that it was possible to observe simultaneously at the assigned roll angle. The full eMPT suite was used to locate the optimal subset of three nearby dithered pointings that provided the best target coverage overall. 
In choosing the final dithered pointings, consideration was also given to the parallel NIRCam exposures achieving good PSF pixel sampling.

In programmes in which the NIRCam imaging was the primary objective (1180 \& 1181) the starting points for NIRSpec were the  pointings dictated by the nominal NIRCam mosaic of each program which was designed to span the various detector gaps in the camera. However, by allowing each of the baseline NIRCam pointings in the mosaic to differ from its nominal position by up to $\pm0.5$ arcsec and each gap-spanning offset to overshoot by up to 1.0 arcsec, enough leeway was given to the NIRSpec pointings to enable the IPA to optimize the final NIRSpec pointings within the permissible range.  

The eMPT software ensured that the Prism spectra of the targets did not overlap on the detectors. The same MSA configuration was used for the gratings, and their longer spectra were allowed to overlap. This was done in the expectation that the prism spectra can be used to  identify lines arising from the actual target rather than spectral overlap. 
Certain high-priority targets were protected from overlaps in the Grating exposures  by closing the shutters in the Grating MSA mask containing lower priority targets whose Grating spectra overlapped with that of the higher priority target.  
In \deephstgs all targets of priority classes 1 through 5 were cleared of overlapping lower priority targets. In all subsequent programmes, this rule was simplified to only the Grating spectra of priority class 1 targets being purged of overlapping spectra.
There were therefore fewer objects observed with the Gratings than observed with Prism, as indicated in Table~\ref{tab:obssummary}.  

We caution the reader that many grating spectra of objects in lower priority classes do feature emission lines from overlapping spectra, and care should be taken to ensure these are not misinterpreted.

\subsubsection{Exposure time distribution}
Figure~\ref{fig:exposure_times} shows the breakdown of exposure times across the full JADES survey. For visual clarity, it is divided into the `Medium' tiers (which include \medhstgs, \medhstgn, \medjwstgs, 
and \medjwstgn; top panel) which have typical exposure times of $t_{\rm exp} \lesssim 10,000$ s (2.8 h) and `Deep' tiers (\deephstgs, \deepjwstgs, and \ultradeepgs; bottom panel) which have typical exposure times of $10,000~{\rm s}\lesssim t_{\rm exp}\lesssim230,000$ s (3 -- 60 h).  We note that some objects have shorter exposure times than a single dither because observations within the nodding pattern were rejected due to background shutters being contaminated by other galaxies, or due to shutters in the slitlet that failed to open. 
We also note that there are 54 targets in the Medium tiers whose multiple exposures add up to $t_{\rm exp} > 15,000$ s, beginning to approach the depths of the Deep tiers.

\subsection{Survey area}

The complicated overlapping nature of the pointings (see Figures~\ref{fig:footprint_gs}, \ref{fig:footprint_gn}) means that the total survey area is non-trivial to compute.  We provide the area within the outline defined by the footprint of useable MSA real-estate of each pointing contributing to a given tier in a given field.  This information is provided in Table~\ref{tab:areas}.

\begin{table}
\begin{center}
\caption{Survey areas probed for each Tier and field.  \medjwstgs is also reported for the early and late stages of the observations, labelled GSa and GSb respectively, which refer to the observations taken pre- and post-September 2023 respectively (see Section~\ref{sec:target_selection} for more details).  \jwst-based area refers to the area with NIRCam coverage at the time of NIRSpec target selection.}
\label{tab:areas}
\begin{tabular}{llcc}
\hline
Field / Tier name &  & area              & \jwst-based \\
                  &  & / arcmin$^2$      & area\\
\hline
\deepjwstgs       &       & \phantom{0}7.056  & 100.0\%\\
\deephstgs        &       & \phantom{0}7.056  & \phantom{00}93.5\%$^\dagger$
\\
\ultradeepgs      &       & \phantom{0}7.056  & 100.0\%\\
\medhstgs         &       & 25.284            & \phantom{00}0.0\%\\
\medhstgn         &       & 24.230            & \phantom{00}0.0\%\\
\medjwstgn        &       & 25.886            & \phantom{0}93.5\%\\
\medjwstgs        & full  & 48.831            & \phantom{0}94.6\%\\
                  & GSa   & 14.186            & 100.0\%\\
                  & GSb   & 44.818            & \phantom{0}94.1\%\\
\hline
\end{tabular}
\end{center}
$^\dagger$ Although the original mask design was performed using \hst-only data, with 0\% \jwst-based area, NIRCam data was available at the stage of mask re-design, and this value reports the percentage NIRCam coverage at that time (see \citetalias{DR1} for more details).
\end{table}

\section{Target selection}
\label{sec:target_selection}

All tiers of the NIRSpec component of the JADES survey employed the multi-object spectroscopy (MOS) mode with NIRSpec. This affords more sensitive spectroscopy than that provided by the NIRCam and NIRISS (Near Infrared Imager and Slitless Spectrograph) grisms, but requires targets to be identified within the field of view prior to observations.  We prioritized catalogues either from \hst or \jwst/NIRCam using tier-specific prioritization schemes, all following the same guiding principle: the rarest, highest-redshift objects were given the highest priority, while lower-redshift galaxies with higher number densities were assigned to lower priority classes.  Some rare classes of lower-redshift objects were promoted to higher classes than assigned to the general galaxy population to increase their chances of being selected, but the final allocations were determined by the eMPT software \citep{bonaventura_empt_2023}. 

The prioritization scheme for \deephst was presented in \citetalias{DR1}, while \citetalias{DR3} first presented the scheme for the \medjwst, \medhst and \ultradeep tiers.  Here, we introduce the scheme for \deepjwst, as well as some changes to the \medjwst prioritization scheme as implemented in the GOODS-S pointings taken after September 2023.  However, for completeness we include the prioritization schemes for all tiers in Tables~\ref{tab:priorities_Deep_JWST}-\ref{tab:priorities_Deep_HST}. In these tables, we also provide the fraction of all available targets within each NIRSpec pointing for which spectra were obtained\footnote{We note that the numbers of available targets noted in Table~\ref{tab:priorities_Deep_HST} for \deephst are smaller than those reported in \citetalias{DR1}.  This is because here we use the smaller projected area of available NIRSpec MSA real-estate that excludes unusable area between MSA quadrants and areas that would produce truncated spectra.  The area used to estimate available targets in \citetalias{DR1} was simply defined by the projection of the outer four corners of the MSA quadrants.} as well as the redshift success rates (see Section~\ref{sec:redshift_success}).

In this section, we first describe the catalogues used for prioritising targets for all post-September 2023 observations (Section \ref{subsec:catalogues}).  We then give an overview of the \deepjwst and \medjwst prioritization schemes (Section \ref{subsec:prioritisation_overview}). Finally, we describe certain criteria in more detail for classes 1-7 in Sections \ref{subsec:classes_1_6} and \ref{subsec:class_7}, such as the criteria used in the high-redshift searches by \cite{Hainline2024} and \cite{Endsley2024}, as well as our own dropout and photometric redshift criteria.

\subsection{Base catalogues for post September 2023 Medium/\jwst and (Ultra-)Deep/\jwst pointings}
\label{subsec:catalogues}

The depth of the NIRCam imaging and its high spatial resolution posed significant challenges to catalogue creation. In particular, our JADES spectroscopy targets galaxies from cosmic noon to within the epoch of reionisation which span a significant range of sizes and morphologies. The final mosaics and catalogues accommodate this diversity and are presented in \cite{Johnson2026} and \cite{Robertson2026}, respectively.
However, the imaging reduction and catalogues available at the time of NIRSpec target selection were evolving as more NIRCam imaging became available, and as data reduction and catalogue detection/deblending methods were being refined. Here we describe the construction of the catalogues used as input to the eMPT for final target selection for the \ultradeep and  \deepjwst pointings,  as well as nine of the \medjwst pointings in GOODS-S.  The NIRCam catalogues available for these selections benefited from several improvements relative to those used for \citetalias{DR1} and \citetalias{DR3}, including improved object-level segmentation and more reliable total flux estimates based on Kron apertures.

The NIRCam base catalogue used to select NIRSpec targets was constructed using deep detection images built from stacked, inverse-variance-weighted signal-to-noise ratio mosaics from available $\lambda>2.5\mu$m images.  Subsequent source deblending employed stacked mosaics constructed from NIRCam filters near $\lambda\approx2\mu$m, allowing for small, irregular galaxies at high-redshift to be isolated.  We also exploited the high-redshift searches of \cite{Hainline2024} and \cite{Endsley2024} that employed catalogues constructed with source deblending that was optimised for identifying small, faint, high-redshift galaxies that were, at the time, occasionally missed (either not detected or not sufficiently deblended from nearby objects) in the main catalogue.  

We first concatenated the \cite{Hainline2024} and \cite{Endsley2024} sources  into a single catalogue, maintaining target centroid and redshift information from the \cite{Hainline2024} catalogue whenever there was a match between the two within 2\arcsec.  This concatenated catalogue supplemented and replaced objects in the base catalogue if there was a match within 2\arcsec.   This hybrid catalogue was then further supplemented with \hst-based targets from the catalogue that was constructed prior to NIRCam imaging in the region, the creation of which is described extensively in \citetalias{DR1} and \citetalias{DR3}.  Objects from the \hst-based catalogue were added if there was no match to the NIRCam catalogue within a radius of 2\arcsec.  This extended the NIRSpec target list to areas beyond the NIRCam coverage at the time, and also allowed it to be supplemented with possible \hst-detected sources that did not meet the detection criteria from the NIRCam-based catalogues.  

For information on the underlying catalogues used to populate NIRSpec target lists in GOODS-S pre-September 2023, and GOODS-N, please see \citetalias{DR1} and \citetalias{DR3}.


As described above, the NIRCam catalogues were constantly improving over the course of the survey.  
For simplicity, Table~\ref{tab:priorities_Medium_JWST} separates the \medjwst target statistics by field (GOODS-S and GOODS-N) and further sub-divides GOODS-S into early (GSa) and late (GSb, post September 2023) observations.  This separation highlights changes to the prioritization strategy of the highest priority targets over time and reflects differences in target number densities due to updates in the underlying base catalogue.  These changes mostly affected photometric redshift quality (see Section~\ref{subsubsec:Medium/JWST}) and total F444W flux estimates for Class 7 (see Section \ref{subsec:class_7}).

\subsection{Prioritization overview}
\label{subsec:prioritisation_overview}

\subsubsection{Deep/JWST priorities}
\label{subsubsec:deep_jwst_prioritisation}

In the \deepjwst portion of the survey, the top priority class is occupied by the highest redshift candidates ($z>10$). The most robust of these, based on visual inspection, were used to optimize the pointing (Class 1), while candidates deemed less robust on visual inspection were placed in Class 1.1 and were assigned shutters if possible after the pointing was set.
One object at $z_{\rm spec} = 3.475$ was placed in Class 1 as, in projection, it lay very close to our best $z\approx14$ candidate and its spectrum was valuable to the analysis of that target (see \citealt{Carniani2024_GS_z14}).  

We chose to centre the pointing on the lower portion of the NIRCam mosaic (see Figure~\ref{fig:footprint_gs}, right panel) which benefitted from very deep imaging, and coverage from multiple medium band filters within the JADES Origins Field \citep[JOF][]{Eisenstein2023_JADES}.  This was following a search of all high-redshift candidates over the field while assessing their predicted NIRSpec PRISM signal-to-noise, and finding several promising candidates within that pointing.

The second and third priority classes were populated with $z>8$ candidates based on their apparent magnitude in a filter sampling the rest-frame UV (see Table~\ref{tab:priorities_Deep_JWST}).  Our primary aim for these classes was redshift identification.

We then turned our focus to $5.7<z<8$ galaxies. We up-weighted candidates with rare properties, such as exceptionally strong emission lines, very blue $\beta$ slopes, candidate AGN and sources with ALMA detections, or strong line emitters identified from MUSE or FRESCO \citep{Maseda2020,Oesch2023}. These unusual objects were placed in priority class 3.1.  This is important because objects like these with such low number densities might never be assigned a shutter otherwise. 
Priority Class 4 was intended to supply spectra with a high S/N in emission lines (H$\alpha [\mathrm{S/N}] \gtrsim 25$ when within the detector wavelength range, H$\beta [\mathrm{S/N}] \gtrsim 10$ otherwise) within $5.7<z<8$, which set the apparent magnitude limit following a conversion of rest-UV absolute magnitude to predicted H$\alpha$ flux using the \citet{Kennicutt2012} star formation rate conversions.  

The fifth priority class was intended to build a small sample of bright galaxies at $z>2$ with high S/N in the continuum across the entire survey but contained no galaxies in \deepjwst.

The sixth priority class captured fainter targets at $z>5.7$ for which we still expect to be able to derive a redshift.  

Priority class seven represents the main bulk of the intermediate redshift galaxy sample for which galaxies were divided into redshift bins between $4.5<z<5.7$ and $1.5<z<4.5$ in bins of $\Delta z=1$.  Class 7 represents a class designed to amass a statistically useful sample across all tiers of the survey.  It was intended for studying galaxy mass assembly from cosmic-noon to the end of the epoch of reionisation. Within \deepjwst, our aim was to target the lower-mass end of galaxies at these redshifts, in contrast to the shallower tiers.

Galaxies at these redshifts identified as likely quiescent, or likely AGN from MIRI/X-ray detections \citep{Lyu2024} are placed first in descending redshift order before assigning shutters to the general galaxy population, which was also ordered by decreasing redshift. However, the low target density of rare targets in classes 7.1-7.4 meant that no targets were assigned to shutters in \deepjwst.  For the remaining classes (7.5-7.9) a simple magnitude limit was set based on the F444W magnitude, the reddest filter, to better approach a mass limited sample.  The lowest redshift bin ($1.5<z<2.5)$ was split into two with the brighter targets being placed first.  We note that the F444W filter gains higher and higher contributions from young stellar populations (or nebular emission) with increasing redshift, leading to a selection function approaching a limit influenced heavily by the current SFR (rather than just stellar mass) in the highest redshift bin.  Additionally, bursty star formation histories at high redshifts can lead to large variations in F444W flux \citep{Simmonds2025, Wang2025}, further complicating the selection as a function of stellar mass.  

Classes eight and nine represent the filler classes, where we target fainter galaxies than captured in higher classes, or lower redshift objects.

The priority class criteria with numbers of targets are presented in Table~\ref{tab:priorities_Deep_JWST}.

\begin{table*}
\caption{Target prioritization categories for Deep/\jwst.}
\centering          
\begin{tabular}{c c c c c c c c c}   
  \hline\hline
Priority   & Redshift        & Criteria                                          & Targets        &  \multicolumn{2}{c}{Success rate}  \\
           &                 &                                                   & (possible targets /MSA) & Success & Interloper \\

\hline                    
1          & $z > 10$         & $m_{\rm UV} < 29.5$ (V.I. Class = 0)$^1$         &  8 (8)         &  57 \% &  0 \%$^{2}$ \\ 
1.1        & $z > 10$        & $m_{\rm UV} < 29.5$ (V.I. Class = 1)$^1$          &  2 (7)        &  50 \% &  0 \%  \\
\\
2          & $z > 8$         & $m_{\rm UV} < 29.5$                               &  13 (65)       &  54 \% &  23 \%  \\
3          & $z > 8$         & $29.5 < m_{\rm UV} < 30.5$ or                     &  1 (30)        &  0 \% &  0 \%  \\
           &                 & $m_{\rm UV} > 3\sigma$                            &                &    &   \\
\\
3.1        & $z>5.7$         & rare targets$^{3}$                                &  2 (8)         &  100 \% &  0 \%  \\
\\
4          & $5.7 < z < 8$   & $m_{\rm UV}< 27.5$                                &  1 (8)         &  100 \% &  0 \%  \\
5          & $z > 2$         & $m_{\rm AB}<22.5$                                 &  0 (1)         &   --- &  --- \\
\\
6.1        & $5.7 < z < 8$   & $27.5<m_{\rm UV}<29$                              &  17 (111)      &  59 \% &  12 \%  \\
6.2        & $5.7 < z < 8.5$ & $29<m_{\rm UV}<30$                                &  11 (69)       &  36 \% &  9 \%  \\
\\
7.1        & $4.5<z<5.7$     & rare targets$^{4}$                                &  0 (0)         & --- &  ---  \\
7.2        & $3.5<z<4.5$     & ''                                                &  0 (1)         & --- &  ---  \\
7.3        & $2.5<z<3.5$     & ''                                                &  0 (2)         & --- &  --- \\
7.4        & $1.5<z<2.5$     & ''                                                &  0 (8)         & --- &  --- \\
7.5        & $4.5<z<5.7$     & $F444W < 27.5$                                    &  3 (36)        & 33 \% &  33 \%  \\
7.6        & $3.5<z<4.5$     & $F444W < 27.5$                                    &  18 (133)      & 89 \% & 5.6 \% \\
7.7        & $2.5<z<3.5$     & $F444W < 27.5$                                    &  26 (264)      & 92 \% & 3.8 \% \\
7.8        & $1.5<z<2.5$     & $F444W < 26.5$                                    &  11 (191)      & 73 \% & 18 \% \\
7.9        & $1.5<z<2.5$     & $26.5<F444W<27.5$                                 &  12 (158)      & 83 \% & 8.3 \% \\
\\
\hline
8.0 \& 8.1 & $z>1.5$         & $F444W<29$\,mag or $S/N({\rm H}\alpha)>20$        &  50 (2053)     & 54 \% & 0 \%  \\
8.2        & $z<1.5$         & $F444W<29$\,mag                                   &  31 (1378)     & 55 \% & 16 \%  \\
8.3        & no redshift cut & $F444W<30$                                        &  24 (2221)     & 33 \% &  ---  \\
9          & ---             &                                                   &  5 (257)       & 20 \% &  ---  \\
\hline
\label{tab:priorities_Deep_JWST}
\end{tabular}
\\
\raggedright
$^{1}${Visual inspection (V.I.) Class = 0 are the most robust candidates, while V.I. Class = 1 are less robust.} \\
$^{2}${We remove one target, 183349, from our success rate calculation. This object is the companion of the z=14.3 galaxy (see \citealt{Carniani2024_GS_z14}). It was known to be low redshift (clearly does not drop out in blue filters), but we assigned it as priority 1 to ensure it got a spectrum since we felt it necessary to robustly study the z=14.3 galaxy. This target has $z=3.476$} \\
$^{3}${Rare targets includes:
L.E.$^{1}$($F_{\rm line}\geq10^{-17.3}$), very blue sources from \citet{Topping2024}, candidate AGN, and sources with ALMA detections.
} \\
$^{4}${Rare targets includes:
L.E.$^{5}$($F_{\rm line}\geq10^{-17.8}$), candidate quiescent galaxies, candidate AGN (including sources with X-ray detections), and sources with ALMA detections.
}\\
$^{5}${Strong line emitters (L.E., units erg\,cm$^{-2}$\,s$^{-1}$) were selected based on measurements from FRESCO or MUSE, or targets with a F410M excess.} \\

\end{table*}

\subsubsection{Medium/\jwst priorities}
\label{subsecsub:medium_jwst_prioritisation}

The prioritization scheme for \medjwst follows that of \deepjwst very closely, only with shallower magnitude limits that take account of the shorter exposure times, as indicated in Table~\ref{tab:priorities_Medium_JWST}. However, the \medjwst pointings were taken over a year baseline and the scheme changed subtly over this time:
\begin{itemize}
  \item The highest priority class has a lower redshift limit (see Section~\ref{subsubsec:high_pri_medjwst} for more specifics as this changed over the course of the survey).
  \item A galaxy over-density ($z\sim7.3$) was identified and included in the top priority class in some pointings.
  \item Some rare galaxies were up-weighted to class 5 (rather than class 3.1 in \deepjwst) that would otherwise have sat in classes 7.1-7.4 to increase the likelihood that they were assigned a shutter.
\end{itemize}

\subsection{Classes 1-6; high-redshifts and rare galaxies}
\label{subsec:classes_1_6}

\subsubsection{Independent searches for high-redshift galaxies}
\label{subsubsec:high_z_searches}

For NIRSpec pointings taken after September 2023 (9 out of 15 \medjwst pointings and \deepjwst\footnote{This is also true for the \ultradeep observations, but we focus here on describing \deepjwst and updated \medjwst prioritization schemes, since \ultradeep was described in \citetalias{DR3}.}), the highest priority classes were primarily populated with sources identified in \cite{Hainline2024} and \cite{Endsley2024}.  \citet{Hainline2024} identify high-redshift ($z>8$) galaxy candidates from photometric redshifts derived using \eazy \citep{Brammer2008} with an updated template set optimized to high-redshift galaxy candidates. Their selection imposes signal-to-noise criteria requiring at least two photometric bands with S/N$>5$, an integrated probability for $z>7$ above 70\%, and, if a low-redshift solution exists, the solution has a $\Delta\chi^2 = \chi^2_{(min,z<7)}-\chi^2_{min}$ greater than 4.  These targets were divided between classes 1 and 3 in \deepjwst\ and 
classes 1b and 3b for \medjwst, based on their apparent magnitude in a filter sampling the rest-frame UV (the brighter of $F150W$ or $F200W$ fluxes). \citet{Endsley2024} employ Lyman-break colour selections to identify high-redshift galaxy samples.  For galaxies at $z\sim 7-9$ they require $F090W-F115W>1.5; F115W-F200W<1.2; F090W-F115W>F115W-F200W+1.5$.  However, we use their photometric redshift estimates, derived using \beagle \citep{Chevallard2016}, to divide the objects between priority classes. 

In \deepjwstgs, $z>10$ candidates from \citet{Hainline2024} and \citet{Endsley2024} were selected for priority Class 1.  These samples were supplemented with additional high-redshift candidates identified from the base NIRCam catalogue, which were assigned to priority Class 2 and lower in \deepjwst, and to the highest priority classes in \medjwst.   For the base catalogue (remaining NIRCam targets in the JADES catalogue not contained within the \citealt{Hainline2024} or \citealt{Endsley2024} high-redshift searches), the sources were either identified from 
a Lyman-break dropout selection (Section~\ref{subsubsec:dropout}), or using photometric redshifts derived from both \eazy \citep{Brammer2008} and \beagle \citep{Chevallard2016} described in Section~\ref{subsubsec:photo_z}. 

\subsubsection{Base catalogue dropout criteria}
\label{subsubsec:dropout}

For the dropout selection we employed permissive criteria with a relatively small drop in flux between detection and dropout bands (which cover wavelengths bluer than the redshifted Lyman-$\alpha$ break which are subject to high IGM absorption), while also considering the spatially varying depth of NIRCam imaging.  The depth varies significantly over the GOODS-S field thanks to the planned two-tier imaging strategy with the deepest region overlapping the \textit{Hubble} Ultra-deep field \citep{Beckwith2006_HUDF}. The NIRCam parallels of the JADES Origins Field and Deep/HST pointings \citep[GO 3215, 1210 parallels][]{Eisenstein2023_JOF} further contribute to the non-uniform depth across GOODS-S.  

To take account of these depth variations, we define break strength compared to the faintest measurable magnitude using the $3\sigma$ limiting depth in the dropout band.  Specifically we apply the following selection criteria:  
\begin{align*}
    {\rm min}[m_{\rm drop},m^{3\sigma}_{\rm lim}] - m_{\rm det}  >& \,0.8 \\
    \,\,m_{\rm det} - m_{\rm det+1} <& \,0.4 \\
\end{align*}
Here, $m_{\rm drop}$ is the apparent magnitude in the dropout filter, $m^{3\sigma}_{\rm lim}$ is the 3$\sigma$ limiting magnitude, and $m_{\rm det},\, m_{\rm det+1}$ are the apparent magnitudes in the first and second filters redward of the Lyman-$\alpha$ break, respectively.  The second colour cut is employed to exclude very red low-redshift galaxy interlopers. We use 0.3 arcsec diameter aperture photometry, which optimises S/N in the colour across the Lyman-break, while roughly approximating the open area of a single MSA shutter. 

The permissive criteria are intended to avoid biasing the sample against redder, intrinsically high-redshift galaxies with modest observable Lyman breaks (relative to the imaging depths), while relying on visual inspection to reject obvious low-redshift contaminants.

F115W (or higher) dropouts ($z\sim9$) were distributed between Classes 2 and 3 based on their apparent magnitude limits, while F090W dropouts ($z\sim7$) were distributed between Classes 4 and 6 (though see Section \ref{subsubsec:high_pri_medjwst} for use of F090W dropouts at higher priority for pointing selection).

\subsubsection{Base catalogue photometric redshift criteria}
\label{subsubsec:photo_z}

When including high-redshift galaxy candidates from photometric redshift estimates we separated objects into `robust' and `possible' categories.  

To be characterised as `robust', an object was required to meet several criteria: 
\begin{itemize}
    \item The \eazy-derived photometric redshift exceeds the limit for the relevant class.
    \item This estimate agrees with the 95\% credible interval of either the primary or secondary peak in the \beagle-derived redshift posterior probability distribution, or the difference in \beagle and \eazy redshift is $\Delta{z}<0.1$.
    \item The integrated posterior probability derived from \beagle exceeds 0.9 for $z>6$ when selecting galaxies at $z>8$ (or $z>4$ when selecting galaxies at $5.7<z<8$).
    \item Objects with redshift quality, as defined by equation (8) of \citet{Brammer2008}, with value $\geq30$ are excluded. This limit rejects imaging artefacts in the catalogue that contaminate high-redshift samples, while retaining plausible high-redshift objects that might not be well-fit by the models. 
\end{itemize}

Any object classed as `possible' from the photometric redshift estimates had an \eazy or \beagle (primary or secondary) photometric redshift solution above the threshold for the given class, but didn't meet the above criteria (excluding the quality cut, which was applied to both categories). 

Objects identified with `robust' photometric redshifts were placed in Class 2 or 4 dependent on their redshift and if they met the magnitude cuts.  Objects identified as `possible' were added to Class 3 or 6 dependent on their redshift, regardless of whether their brightness would have qualified them to occupy a higher priority class. 

\subsubsection{Assembly of high-redshift classes}

For high-redshift objects placed into Classes 1-6 (i.e. $z\gtrsim5.7$, rather than rare lower-redshift galaxies) the redshift information was drawn from one of the three sources described above (high-redshift galaxy searches, dropout criteria or photometric redshifts in Sections \ref{subsubsec:high_z_searches}-\ref{subsubsec:photo_z}).  In principle, multiple redshift estimates are available for each object.  In practice, individual classes were assembled in the following order: (i) independent searches for high-redshift galaxies (with precedence given to the redshift information from \citealt{Hainline2024}); (ii) dropout candidates from the base catalogue not captured by (i); and (iii) photometric redshift estimates meeting the class redshift boundaries but not previously included in (i) and (ii).  

The redshift source used for prioritization, together with the \beagle and \eazy photometric redshift information (where applicable), is listed in Table~\ref{tab:prioritisation_redshift_information}.

\subsubsection{Changes to highest priority targets in Medium/\jwst}
\label{subsubsec:high_pri_medjwst}

Figure~\ref{fig:GS_detail} shows all \medjwst pointings overlaid on the NIRCam mosaic, with the GSa and GSb pointings shown separately.  The objects in the top two priority classes are over-plotted, and those that were targetted are further highlighted. Although our top priority class required $z>9$ in the GSa pointings, some of our highest priority targets included objects within an over-density at $z\sim7.3$ 
identified via excess flux in the F410M filter.
Those targetted are shown as X in the figure, and separated out as priorities 1X and 2X in Table~\ref{tab:priorities_Medium_JWST}.

We did not re-observe targets in successive visits to the same region, though overlap was allowed between pointings in the same visit. Therefore, objects targetted in GSa were not used for pointing optimisation in GSb. Due to this constraint and the relatively high pointing density in GSb, we adjusted the redshift criteria from $z>9$ (in GSa) to $z>8$ to maintain sufficient target density for pointing optimization with the IPA.  Additionally, classes 1 and 2 were both used for the pointing optimization for GSb pointings (vs. class 1 used in other tiers and GSa).  

Figure~\ref{fig:GS_detail} also illustrates a paucity of targets at $z>8$ that met our brightness threshold (see Table~\ref{tab:priorities_Medium_JWST}) in the north-west portion of the mosaic.  We therefore supplemented the top priority classes with galaxies selected as $F090W$ dropouts, corresponding to redshifts as low as $z\gtrsim6.5$.  These objects are distinguished in Figure~\ref{fig:GS_detail} as indicated in the legend.

\subsubsection{Visual inspection}
\label{subsubsec:visual_inspection}

All objects identified as class 1-6 were visually inspected to reject obvious low-redshift objects.  Objects deemed to not be robust high-redshift candidates, but still possible, were generally demoted to class 3 if $z>8$ and class 6 if $z>5.7$.  Any objects selected from dropout criteria that were deemed more likely to be at lower redshift were distributed into lower classes based on their photometric redshift estimates and apparent magnitudes.

\subsection{Class 7; intermediate redshifts}
\label{subsec:class_7}

The redshift bins for class 7 in \deepjwst and \medjwst (see Tables~\ref{tab:priorities_Medium_JWST} and \ref{tab:priorities_Deep_JWST}) were allocated using information from both \eazy and \beagle photometric redshifts.  The redshifts are required to be `robust' following the criteria described in Section~\ref{subsubsec:photo_z} (without any constraints on integrated probability). 
Among the different failure modes for photometric redshift estimation is the confusion of prominent breaks, such as the Lyman and Balmer breaks.  The comparison between primary and secondary peaks in the \beagle posterior allows for possible break confusion by \beagle.  However, there is a category of objects that might be missing from their correct redshift bin if \eazy had incorrectly assigned the break. This is a source of possible incompleteness or bias in these samples but has been employed to ensure each object is only assigned to a single redshift bin.    

Within the redshift slices of Class 7, some rare or unusual galaxies were up-weighted in their class to improve the chances of being allocated a shutter in the MSA design. Specifically, in Class 7, passive galaxies or galaxies with AGN signatures such as associated X-ray detection were placed first in descending redshift slices before the `normal' galaxy population, again in descending redshift order of priority. 
Passive galaxies were identified from the redshift-dependent UVJ selection of \citet{Leja2019}.
However, some passive galaxies were selected independently and up-weighted to improve targetting success.

\begin{figure*}
    \centering
    \includegraphics[width=0.9\linewidth]{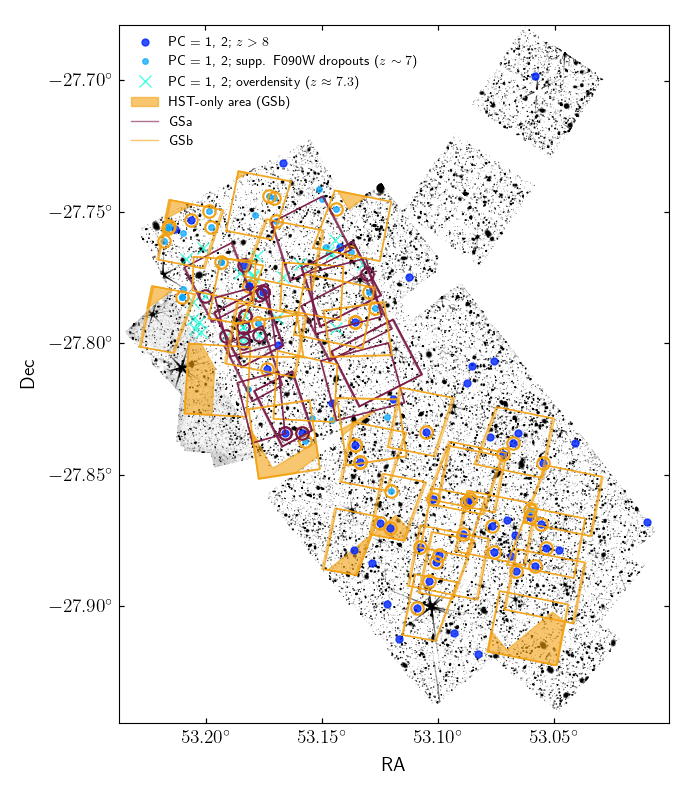}
    \caption{Layout of Medium/\jwst pointings in GOODS-S across the two main epochs, GSa (purple; prior to Sep 2023) and GSb (orange; after Sep 2023). Shaded regions within MSA footprints show areas where NIRCam imaging was not available when target selection was performed for the observation in question.
    Dark blue points show our nomimal priority classes (PC) = 1\&2 targets (with $z>8$), used to set the pointing centres, while those with a coloured circle around them were actually observed. After the original pointings in GSa, there were very few unobserved PC=1 \& 2 targets in the northern part of the field, leading us to supplement these with F090W-dropout selected galaxies, which have $z\gtrsim6.5$ (Class 2d in Table~\ref{tab:priorities_Medium_JWST}; light blue circles).  In addition, candidate members of a $z\approx7.3$ overdensity, identified from medium-band excesses (Class 1x in Table~\ref{tab:priorities_Medium_JWST}; teal crosses) were added to the top priority classes in some GSa pointings.
    }
    \label{fig:GS_detail}
\end{figure*}

\begin{landscape}
\begin{table}
\caption{Target prioritization categories for Medium/\jwst.
 The categories for early (pre September 2023) and late (post September 2023) GOODS-S pointings are reported in columns labelled GSa and GSb due to their differing top priority class structure, and different base catalogues used for target selection and target density estimates.  The total area with \jwst-only pre-selection is 100\% in GSa, 94.0\% in GSb and 93.5\% in GN, where GSa covers. }       
\begin{tabular}{c c c c c c c c c c c c }   
  \hline\hline
Priority   & Redshift        & Criteria (if \jwst-based )                         & Criteria (if \hst-based)             & \multicolumn{3}{c}{Unique allocated targets} & \multicolumn{2}{c}{Success rate}  \\   
           &                 &                                                   &                                     & \multicolumn{3}{c}{(possible targets /MSA coverage)} & & \\
           &                 &                                                   &                                     & GSa        & GSb        & GN       & Success & Interloper \\
\hline  
1a         & $z>9$           & $m_{\rm UV} < 27.9$ (V.I. Class = 0)$^1$          &                                     & 3 (6)      &            & 12 (13)  & 53~\% & 33~\%  \\  
1b         & $z>8$           & Hainline, Endsley, $m_{\rm UV} < 28$              &                                     &            & 29 (32)    &          & 76~\% & 21~\% \\
\\
1x         & $z\approx7.3$   & Overdensity                                       &                                     &   3        &            &          & 100~\% & 0~\%  \\
\\
2a         & $z > 9$         & $m_{\rm UV} < 27.9$ (V.I. Class = 1$^2$)              &                                     & 1 (9)      &            & 0 (0)    & 0~\% & 0~\%  \\   
2c         & $z>8$           & $m_{\rm UV} < 28$                                 &                                     &            & 5 (11)     &          & 20~\% & 40~\% \\
2d         & $z>6.5$         & $m_{\rm UV} < 28$                                 &                                     &            & 15 (25)    &          & 47~\% & 20~\% \\ 
2x         & $z\approx7.3$   & Overdensity                                       &                                     &   5        &            &          & 100~\% & 0~\%  \\
\\
3a         & $z > 9$         & $27.9\le m_{\rm UV} <28.5$ or                     &                                     & 1 (9)      &            & 7 (15)   & 12~\% & 38~\%  \\
           &                 & oddballs                                          &                                     &            &            & 2        & \\
3b         & $z > 8$         & Hainline, Endsley $28.5\le m_{\rm UV} < 29.8$     &                                     &            & 8 (29)     &          & 75~\% & 0~\%      \\
3c         & $z > 8^3$       & $28.0\le m_{\rm UV} < 28.5$                       & $28.0<F160W < 28.5$                 &            & 8 (31)     &          & 12~\% & 38~\%        \\
3d         & $z>6.5$         & $28.0\le m_{\rm UV} < 28.5$                       &                                     &            & 9 (37)     &          & 22~\% & 11~\%      \\
\\
4          & $5.7 < z < 8$   & $m_{\rm UV}< 26.5$ or                             &                                     & 7 (20)     & 5 (22)    & 5 (12)  & 88 \% & 6~\%   \\
           &                 & oddballs                                          &                                     & 2           & 1         & 6  \\
\\
5          & $z > 2^4$       & $m_{\rm AB}<22$ \& oddballs                       & $F160W<22$                          & 1 (10)     & 9 (47)     & 10 (41)  & 85 \% & 5 \%   \\
\\
6.1  & $5.7 < z < 9$  & $26.5<m_{\rm UV}<28$ or                           & $F160W<28$                          & 13 (72)    & 50 (202)  & 57 (185) & 60 \% & 10 \%  \\ 
           &                 & oddballs                                          &                                     & 4          & 8         & 2   \\ 
6.2a       & $5.7<z<6.5$     & $F444W < 27$                                      &                                     & 3 (22)     &           & 11 (54) & 21 \% & 36 \% \\ 
6.2b       & $5.7 < z < 8.5$ & $28<m_{\rm UV}<28.5$ or                           & $F160W>28$                          &            & 38 (214)  &   & 47 \% & 18 \%  \\
\\
7.1        & $4.5<z<5.7$     & UVJ  \& $F444W < 27$; X-ray sources               & $F160W < 28$                        & 0 (1)      & 0 (9)      & 3 (13)    & 33 \% & 67 \% \\
7.2        & $3.5<z<4.5$     & UVJ  \& $F444W < 27$; X-ray sources               & $F160W < 28$                        & 0 (6)      & 1 (5)      & 5 (12)   & 100 \% & 0 \% \\
7.3        & $2.5<z<3.5$     & UVJ  \& $F444W < 27$; X-ray sources               & $F160W < 27.5$                      & 5 (18)     & 5 (35)     & 11 (30)  & 90 \% & 5 \%  \\
7.4        & $1.5<z<2.5$     & UVJ  \& $F444W < 27$; X-ray sources               & $F160W < 27.5$                      & 9 (33)     & 21 (68)    & 9 (25)   & 87 \% & 8 \%  \\
7.5        & $4.5<z<5.7$     & $F444W < 27$  or                                  & $F160W < 28$                        & 15 (76)    & 40 (214)   & 54 (297) & 79 \% & 10 \%  \\
           &                 & oddballs                                          &                                     &            &            & 4\\
7.6        & $3.5<z<4.5$     & $F444W < 27$                                      & $F160W < 28$                        & 44 (226)   & 109 (702)  & 61 (433)  & 73 \% & 8 \%  \\
7.7        & $2.5<z<3.5$     & $F444W < 27$                                      & $F160W < 27.5$                      & 68 (446)   & 188 (1394) & 121 (888) & 82 \% & 7 \%  \\
7.8        & $1.5<z<2.5$     & $F444W < 26$                                      & $F160W < 27.5$                      & 58 (440)   & 148 (1335) & 93 (913)  & 83 \% & 7 \%  \\
7.9        & $1.5<z<2.5$     & $26<F444W<27$; or                                 &                                     & 9 (189)    & 110 (1042) & 43 (615)  & 62 \% & 9 \%  \\
           &                 &  oddballs                                         &                                     &            &   1        & 3    \\ 
\\
\hline
8.0 \& 8.1 & $z>1.5$         & $F444W<28$\,mag or $S/N({\rm H}\alpha)>20$        & $F160W>28.5$                        & 211 (4625) & 298 (6771) & 139 (3191) & 39 \% & 1.5 \%   \\
           &                 &                                                   & \& has GAIA2 coords \\
8.2        & $z<1.5$         & $F444W<28$\,mag                                   & $24.5<F160W<29$                     & 30 (1856)  & 159 (4994) & 156 (3617) & 38 \% & 12 \% \\
           &                 &                                                   & \& has GAIA2 coords \\
8.3        & no redshift cut & $F444W<29$                                        & $F160W>29$                          &            & 172 (7854) & 107 (3294)  & 21 \% & ---  \\
           &                 &                                                   & \& has GAIA2 coords \\
9          & ---             &                                                   & $F160W>24.5$                        &  21 (1314) & 44 (2452)  & 29 (1235)  & 15 \% & ---  \\
\hline
\label{tab:priorities_Medium_JWST}
\end{tabular}
\\
$^{1}${Visual inspection (V.I.) Class=0 are the most robust candidates.}
$^{2}${V.I. Class=1 are deemed less robust but still plausible.}
$^{3}${Redshift cut was higher ($z>8.5$) for \hst-based selections.}
$^{4}${Additional redshift upper limit ($z<5.7$) was employed for \hst-based selections.}
\end{table}
\end{landscape}

\begin{table}
    \centering
    \begin{tabular}{| c | c c c | c c c |}
    \hline
        & \multicolumn{3}{c|}{GSa} & \multicolumn{3}{c|}{GSb} \\
        \hline
       Priority  & No. & Succ. & Interlop. & No. & Succ. & Interlop. \\
       \hline
        7.5  & 15 & 67~\% & 20~\% & 40 & 78~\% & 10~\% \\
        7.6  & 44 & 64~\% & 18~\% & 110 & 80~\% & 4~\% \\
        7.7  & 68 & 82~\% & 12~\% & 188 & 84~\% & 6~\% \\
        7.8  & 58 & 72~\% & 7~\% & 148 & 88~\% & 5~\% \\
        \hline
    \end{tabular}
    \caption{Comparative success rates between `GSa' (prior to September 2023) and `GSb' (after September 2023) from Medium/\jwst in GOODS-S. See Table~\ref{tab:priorities_Medium_JWST} for priority class descriptions.}
    \label{tab:gsa_gsb_compare}
\end{table}

\begin{table*}
 \caption{Target prioritization categories for 3215 `UltraDeep'} 
\centering          
\begin{tabular}{c c c c c c }   
  \hline\hline
Priority & Redshift    & Criteria                               & Targets    & \multicolumn{2}{c}{Success rates}\\
         &             &                                        & (possible targets /MSA) & Success & Interloper \\
\hline                    
1.1      & $z>11$      & $m_{AB}<30$                            & 4 (4) & 100 \%$^{3}$ & --- \\
1.2      & $z>11$      & $m_{AB}<30$ and less reliable phot-$z$ & 0 (1) & --- & ---\\
2.1      & $10<z<11$   & $m_{AB}<30$                            & 0 (4) & --- & --- \\
2.3      & $8<z<10$    & $m_{AB}<30$                            & 6 (18) & 83 \% & 0 \% \\
2.4      & $8<z<10$    & $m_{AB}<30$ and less reliable phot-$z$ & 2 (6) & 0 \% & 0 \% \\
\\
3.1      &             & rare objects$^{1}$                     & 5 (22) & 60 \% & 0 \% \\
3.2      &             & rare objects$^{2}$                     & 4 (21) & 75 \% & 0 \% \\
\\
4.1      & $7<z<8$     & $m_{AB}<30$ from Endsley et al.         \\
         & $5.7<z<8$   & $m_{AB}<28.5$ from other phot-$z$      & 3 (50) & 100 \% & 0 \% \\
4.2      & $5.7<z<7$   & $m_{AB}<30$ from Endsley et al.        & 8 (44) & 25 \% & 63 \% \\
\\
5.1      & $4<z<5.7$   & $m_{AB} < 28$                          & 15 (145) & 60 \% & 20 \% \\
5.2      & $4<z<5.7$   & $m_{AB} < 29$                          & 23 (250) & 56 \% & 22 \% \\
6.1      & $5.7<z<8$   & $28.5<m_{AB} < 30$                     & 14 (173) & 0 \% & 57 \% \\
6.2      & $4<z<5.7$   & $m_{AB} < 30$                          & 29 (694) & 24 \% & 38 \% \\
7.1      & $2.5<z<4$   & $25<m_{AB}<28$                         & 15 (305) & 93 \% & 0 \% \\ 
7.2      & $2.5<z<4$   & $28<m_{AB}<29$                         & 12 (294) & 67 \% & 0 \% \\ 
7.3      & $1.5<z<2.5$ & $25<m_{AB}<28$                         & 14 (446) & 86 \% & 0 \% \\ 
7.4      & $1.5<z<2.5$ & $28<m_{AB}<29$                         & 15 (426) & 40 \% & 0 \% \\ 
7.5      & $z>1.5$     & $29<m_{AB}<30$                         & 17 (962) & 24 \% & 0 \% \\ 
8.1      & $z<1.5$     & $25<m_{AB}<28$                         & 21 (931) & 48 \% & 9.5 \% \\ 
8.2      & $z<1.5$     & $28<m_{AB}<29$                         & 7 (461) & 14 \% & 29 \% \\ 
8.3      & $z<1.5$     & $29<m_{AB}<30$                         & 5 (612) & 0 \% & 80 \% \\ 
9        &             & class 9 objects in \deephst            & 9 (648) & 11 \% & --- \\
  \hline
\label{tab:priorities_3215}
\end{tabular}
\\
\raggedright
$^{1}${Rare objects includes: blue UV slopes, AGN $7<z<12$, MIRI $z>7$, X-ray $z>4$, medium-band $\log(\rm{line~flux/erg\,cm^{-2}\,s^{-1}})>-18.3$}
\\
$^{2}${Rare objects includes: ALMA, MIRI $z<7$, AGN $4<z<7$, medium-band $\log({\rm line~flux/erg\,cm^{-2}\,s^{-1}})<-18.3$}\\
$^{3}${Class 1.1 targeted four objects from 1210 which had known redshifts.}
\end{table*}

\begin{table*}
\caption{Target prioritization categories for Medium/\hst}  
\centering          
\begin{tabular}{c c c c c c c c c c c}   
  \hline\hline
Priority & Redshift        & Criteria                                      & \multicolumn{2}{c}{Unique allocated targets}         & \multicolumn{2}{c}{Success rate} \\
         &                 &                                               & \multicolumn{2}{c}{(possible targets /MSA coverage)} &     &      \\
         &                 &                                               & GS         & GN  & Success & Interloper \\
\hline                    
  1      & $z>5.7$	     & $F160W < 27.5$; V.I. Class$^{1}$ 0	         & 23 (40)    & 25 (30)                            & 90 \%  &  6 \%  \\
  2.0    & $z>5.7$	     & $F160W < 27.5$; V.I. Class 1                  & \\
         & 	              & $27.5 < F160W < 29$; V.I. Classes 0, 1        & 48 (135)   & 16 (40)   & 61 \% & 11 \%                              \\
  \\
  3.0    & $1.5 < z < 5.7$ & Rare target (e.g., Qui., AGN, ALMA\dots) & 8 (23)    & 12 (20)    & 85 \% & 5 \%                             \\ 
  3.5    & $1.5 < z < 5.7$ & $F160W < 23.5$                                & 11 (122)   & 26 (106)         & 100 \% & 0 \%                       \\
  \\
  4.1    & $4.5<z<5.7$	 & $F160W < 25.5$	                             & 6 (23)    & 3 (16)              & 33 \% & 33 \%     \\	
  5.1    & $4.5<z<5.7$	 & $F160W < 27$                                  & 33 (116)   & 41 (97)            & 55 \%  & 10 \%    \\ 
  6.1    & $4.5<z<5.7$	 & $S/N(\Halpha) > 15$                           & 22 (85)   & 5 (16)      & 26 \%  & 22 \%   \\ 
\\
  4.2    & $3.5<z<4.5$	 & $F160W < 25.5$                                & 11 (57)    & 16 (37)            & 48 \%  & 19 \%      \\ 
  5.2    & $3.5<z<4.5$     & $F160W < 27$                                  & 51 (236)   & 54 (154)           & 64 \% & 10 \%   \\ 
  6.2    & $3.5<z<4.5$     & $S/N(\Halpha) > 15$                           & 23 (162)   & 10 (24)     & 55 \% & 9 \%    \\ 
\\
  4.3    & $2.5<z<3.5$	 & $F160W < 25.5$                                & 45 (248)   & 37 (137)         & 80 \% & 10 \%     \\
  5.3    & $2.5<z<3.5$	 & $F160W < 27$                                  & 54 (481)  & 65 (282)          & 70 \% & 9 \%    \\
  6.3    & $2.5<z<3.5$	 & $S/N(\Halpha) > 15$                           & 22 (240)   & 9 (39)    & 55 \% & 10 \%    \\ 
\\
  4.4    & $1.5<z<2.5$     & $F160W < 25.5$                                & 41 (427)   & 82 (370)         & 87 \% & 2.4 \%    \\
  5.4    & $1.5<z<2.5$	 & $F160W < 27$                                  & 63 (698)  & 98 (603)          & 58 \% & 8 \%     \\ 
  6.4    & $1.5<z<2.5$	 & $S/N(\Halpha) > 15$                           & 21 (395)  & 4 (62)     & 24 \% & 8 \%   \\ 
\\
  7      & $z > 1.5$	     & Has GAIA2 coords and $F160W > 23.5$           & 76 (3504) & 72 (814)          & 39 \%  & 1.4 \%    \\
  7.5    & $z < 1.5$       & Has GAIA2 coords	$23.5 < F160W< 27$           & 77 (2224) & 165 (1679)        & 34 \%  & 11 \%    \\
  7.6    & $z < 1.5$	     & Has GAIA2 coords	$F160W > 27$                 & 26 (1533)  & 27 (496)         & 9 \% & 9 \%   \\ 
  8      & any $z$         & Anything else with $F160W > 23.5$             & 16 (1947)  & 86 (1075)        & 51 \% & ---  \\ 
\hline
\label{tab:priorities_Medium_HST}
\end{tabular}
\\
$^{1}${Targets were assigned one of the following visual inspection (V.I.) classes: (0) Most compelling, (1) Plausible $z>5.7$, but less compelling, (2) Real object but likely $z<5.7$, (-1) Reject.}
\end{table*}

\begin{table*}
\caption{Target prioritization categories for PID 1210 Deep/\hst. The \hst and \jwst entries for each class denote the different priority criteria whether the source was primarily selected from \jwst or \hst (see text for details).  The number of targets per MSA footprint were estimated from the full 3\farcm6 $\times$ 3\farcm4 field of view.}            
\label{table:1}      
\centering          
\begin{tabular}{c c c c c c c }     
\hline\hline       
Priority & Redshift       & Criteria                                                    & Targets         & Success\footnotemark[4] & interloper \\ 
         &                &                                                             & (possible /MSA) & rate                    & fraction \\
\hline                    
1        & $z > 8.5$      & F160W $< 29$ (\textit{HST})\\
         &                & $m_\textrm{UV} < 29.5$ (\textit{JWST})                      & 6 (6)     & 83\%          & 0\% \\
         &                & V.I. Class 0 \\
\hline
2        & $z > 8.5$      & F160W $< 29$ (\textit{HST})\\
         &                & $m_\textrm{UV} < 29.5$ (\textit{JWST})                      & 2 (3)     & 50\%          & 0\% \\
         &                & V.I. Class 1 \\
\hline
         & $z > 8.5$      & F160W $> 29$ (\textit{HST}) \\
3        &                & $30.5 < m_\textrm{UV} < 29.5$ {\textit(JWST)}                        & 3 (4)     & $>$33\%$^{a}$ & $<$33\%$^{a}$\\\\
         &                & V.I. Class 1,0 \\
\hline
        & $5.7 < z < 8.5$ & $M_\textrm{UV} < ?$ (\textit{HST}) \\
4       & $6 < z < 8.5$   & $m_\textrm{UV} < 27.5$ (\textit{JWST})                      & 20 (44)   & 80\%          & 5\%$^b$ \\
\hline 
        & $2 < z <5.7$    & F160W $< 23$ (\textit{HST})\\
5       & $z > 2$         & any filter $< 22.5$ (\textit{JWST})                         & 5 (14)    & 100 \% & 0 \%\\
\hline
        & $5.7<z<8.5$     & F160W $< 29$ (\textit{HST}) \\
6.1     & $5.7<z<8.5$     & (F105W $< 29 |$ F150W $< 29$) (\textit{JWST})               & 9 (57)    & 89\%          & 11\% \\
\\
        & $5.7 < z< 8.5$  & F160W $> 29$ (\textit{HST}) \\
6.2     & $5.7 < z < 6.5$ & F444W\footnotemark[2] $< 27.5$ (\textit{JWST})              & 7 (53)    & 57\%          & 29\% \\
\hline
7.1     & $4.5\leq z<5.7$ & F160W $<29$ (\textit{HST}); F444W\footnotemark[2] $<27.5$ (\textit{JWST})               & 1 (2)     & 0 \% & 0 \% \\
        &                 & UVJ; X-ray sources\\
\\
7.2     & $3.5\leq z<4.5$ & F160W $< 29$ (\textit{HST}); F444W\footnotemark[2] $<27.5$ (\textit{JWST})               & 1 (4)     & 100 \% & 0 \% \\
        &                 & UVJ; X-ray sources\\
\\
7.3     & $2.5\leq z<3.5$ & F160W $<29$ (\textit{HST}); F444W\footnotemark[2] $<27.5$ (\textit{JWST})               & 0 (9)     & --- & --- \\
        &                 & UVJ; X-ray sources\\
\\
7.4     & $1.5\leq z<2.5$ & F160W $<29$ (\textit{HST}); F444W\footnotemark[2] $<27.5$ (\textit{JWST})               & 1 (7)     & 100 \% & 0 \% \\
        &                 & UVJ; X-ray sources\\
\\
7.5     & $4.5\leq z<5.7$ & F160W $<29$ (\textit{HST}); F444W\footnotemark[2] $<27.5$ (\textit{JWST})               & 23 (139)  & 78\%          & 9\% \\
\\
7.6     & $3.5\leq z<4.5$ & F160W $<29$ (\textit{HST}); F444W\footnotemark[2] $<27.5$ (\textit{JWST})               & 31 (298)  & 74\%          & 7\% \\
\\
7.7     & $2.5\leq z<3.5$ & F160W $<29$ (\textit{HST}); F444W\footnotemark[2] $<27.5$ (\textit{JWST})               & 45 (540)  & 78\%          & 7\% \\
\\
7.8     & $1.5\leq z<2.5$ & F160W $<29$ (\textit{HST}); F444W\footnotemark[2] $<27.5$ (\textit{JWST})               & 47 (816)  & 75\%          & 6\% \\
\hline
8.1     & $1.5\leq z<5.7$ & F160W $ > 28.5$,  AND has GAIA2 coords (\textit{HST})       & 20 (1017) & 45 \% & 0 \% \\ 
        &                 & $27.5 <$ F444W\footnotemark[2] $<29$ (\textit{JWST})\\
\\
8.2     & $z<1.5$         & $24.5 < $F160W$ < 29$,  AND has GAIA2 coords (\textit{HST}) & 17 (449)  & 47 \% & 24 \% \\
        &                 & F444W\footnotemark[2] $ < 29$ (\textit{JWST})\\
\\
8.3     & $z<1.5$         & F160W $>29$,  AND has GAIA2 coords (\textit{HST})           & 3 (165)   & 0 \% & 0 \% \\\\
    \hline
9       &                 & fillers (not deliberately rejected)                         & 12 (812)  & 8.3 \% & --- \\
\hline
\label{tab:priorities_Deep_HST}
\end{tabular}

\raggedright
\footnotemark[2]{Denotes photometry derived from Kron apertures.}
\footnotemark[4]{The success rate is the fraction of galaxies targeted who had a spectroscopic redshift measured within $\Delta z=0.1$ of the predicted redshift interval for that priority class. Galaxies lying outside this range are classed as interlopers.}

$^{a}${The spectrum of object 9992 is Class 3 is ambiguous and may show two sources, a low-redshift galaxy at $z=1.962$ and hints of a second galaxy at $z>9$}
$^b${One target in Class 4 for which we did not get a good spectrum, 10035328, is a star (with a proper motion of 0\farcs16 between 
\hst/WFC3 and NIRCam) and we class it as an interloper.}

\end{table*}

\section{Target selection success}
\label{sec:redshift_success}

\begin{table*}
    \caption{Table of spectroscopic redshift obtained from the JADES survey. For brevity, only the top priority classes from Deep/\jwst are shown here, demonstrating the form and content. The full table of over 5000 spectroscopic targets will be available in machine-readable form once published, for now the information is maintained in the \href{https://jades-survey.github.io/scientists/data.html}{released catalogues within the data release}.}
    \centering
    \begin{tabular}{cccccccccccc}
    \hline
       &             &    &     &          &                &      & \multicolumn{2}{c}{Intra-shutter} & \multicolumn{2}{c}{Exposure} \\
      &             &    &     &          &                &      & \multicolumn{2}{c}{offset ($\mathrm{{}^{\prime\prime}}$)} &\multicolumn{2}{c}{ time (h) } \\
\hline
Tier & NIRSpec\_ID & R.A. & Decl. & Priority & $z_{\rm spec}$ & flag & x & y & Prism & Gratings\\
\hline
\deepjwstgs & 183348 & 53.0829371 & -27.8556321 & 1 & 14.1796 & C & -0.06 & 0.071 & 18.5 & 4.6  \\
\deepjwstgs & 183349 & 53.0830522 & -27.8556219 & 1 & 3.4754 & A & 0.128 & 0.063 & 9.2 & 2.3  \\
\deepjwstgs & 20012702 & 53.0600576 & -27.8914295 & 1 & 6.8 & E & 0.073 & 0.006 & 27.7 & 6.9  \\
\deepjwstgs & 20013731 & 53.0647541 & -27.8902378 & 1 & 13.01 & C & 0.068 & 0.01 & 27.7 & 6.9  \\
\deepjwstgs & 20018044 & 53.0742705 & -27.8859192 & 1 & 13.86 & C & -0.046 & 0.049 & 27.7 & 6.9  \\
\deepjwstgs & 20055733 & 53.1076259 & -27.8601385 & 1 & --- & E & 0.03 & -0.096 & 27.7 & 6.9  \\
\deepjwstgs & 20064312 & 53.0724799 & -27.8553518 & 1 & 10.48 & D & -0.065 & 0.003 & 27.7 & 6.9  \\
\deepjwstgs & 20176151 & 53.0707635 & -27.8654376 & 1 & 10.4075 & C & -0.022 & 0.098 & 27.7 & 6.9  \\
\deepjwstgs & 20015720 & 53.1176273 & -27.8881759 & 1.1 & 11.275 & C & 0.025 & 0.112 & 27.7 & 6.9  \\
\deepjwstgs & 20177294 & 53.079003 & -27.8635899 & 1.1 & 10.8 & D & -0.018 & -0.121 & 27.7 & 6.9  \\
\deepjwstgs & 9442 & 53.0696772 & -27.8958958 & 2 & 1.462 & D & -0.061 & -0.092 & 27.7 & 6.9  \\
\deepjwstgs & 12326 & 53.1056073 & -27.8918615 & 2 & 7.9494 & C & 0.043 & 0.056 & 27.7 & 0.0  \\
\deepjwstgs & 17909 & 53.0692252 & -27.8860352 & 2 & 2.0721 & C & 0.008 & 0.021 & 27.7 & 6.9  \\
\deepjwstgs & 20005936 & 53.1199136 & -27.9015789 & 2 & 7.9489 & A & -0.015 & -0.097 & 27.7 & 6.9  \\
\deepjwstgs & 20006347 & 53.1089999 & -27.9008416 & 2 & 8.7142 & A & 0.078 & -0.0 & 27.7 & 4.6  \\
\deepjwstgs & 20015285 & 53.0790506 & -27.8885972 & 2 & 10.2615 & D & 0.014 & 0.018 & 27.7 & 6.9  \\
\deepjwstgs & 20021387 & 53.0671523 & -27.8831711 & 2 & --- & E & -0.015 & -0.049 & 27.7 & 6.9  \\
\deepjwstgs & 20050575 & 53.0703713 & -27.8630784 & 2 & 8.4143 & A & 0.001 & -0.07 & 27.7 & 6.9  \\
\deepjwstgs & 20051718 & 53.1137767 & -27.8623769 & 2 & 7.948 & B & 0.04 & 0.088 & 18.5 & 0.0  \\
\deepjwstgs & 20062446 & 53.1200125 & -27.856452 & 2 & 7.649 & A & -0.032 & -0.002 & 27.7 & 6.9  \\
\deepjwstgs & 20074794 & 53.0671723 & -27.8472697 & 2 & 8.5894 & C & 0.021 & 0.098 & 27.7 & 6.9  \\
\deepjwstgs & 20179485 & 53.0873786 & -27.8603251 & 2 & 7.9539 & A & 0.002 & -0.171 & 18.5 & 4.6  \\
\deepjwstgs & 70378948 & 53.0633829 & -27.850607 & 2 & 2.0791 & C & -0.003 & -0.145 & 9.2 & 2.3  \\
\hline
\label{tab:redshift_table}
\end{tabular}
\end{table*}

\subsection{Data processing}
\label{sec:data_processing}

All spectra included in this release were reduced by the latest version of the NIRSpec GTO pipeline, with updated NIRSpec calibration files.  This includes spectra that were previously presented in \citetalias{DR1} and \citetalias{DR3}, which have been re-reduced for consistency in this data release.  For full details of the data reduction and calibration we refer the reader to \citetalias{DR4_paper2} which also presents the emission line measurements from the spectroscopic dataset.

\subsection{Redshift measurement}

\begin{figure*}
    \centering
    \includegraphics[width=0.99\linewidth]{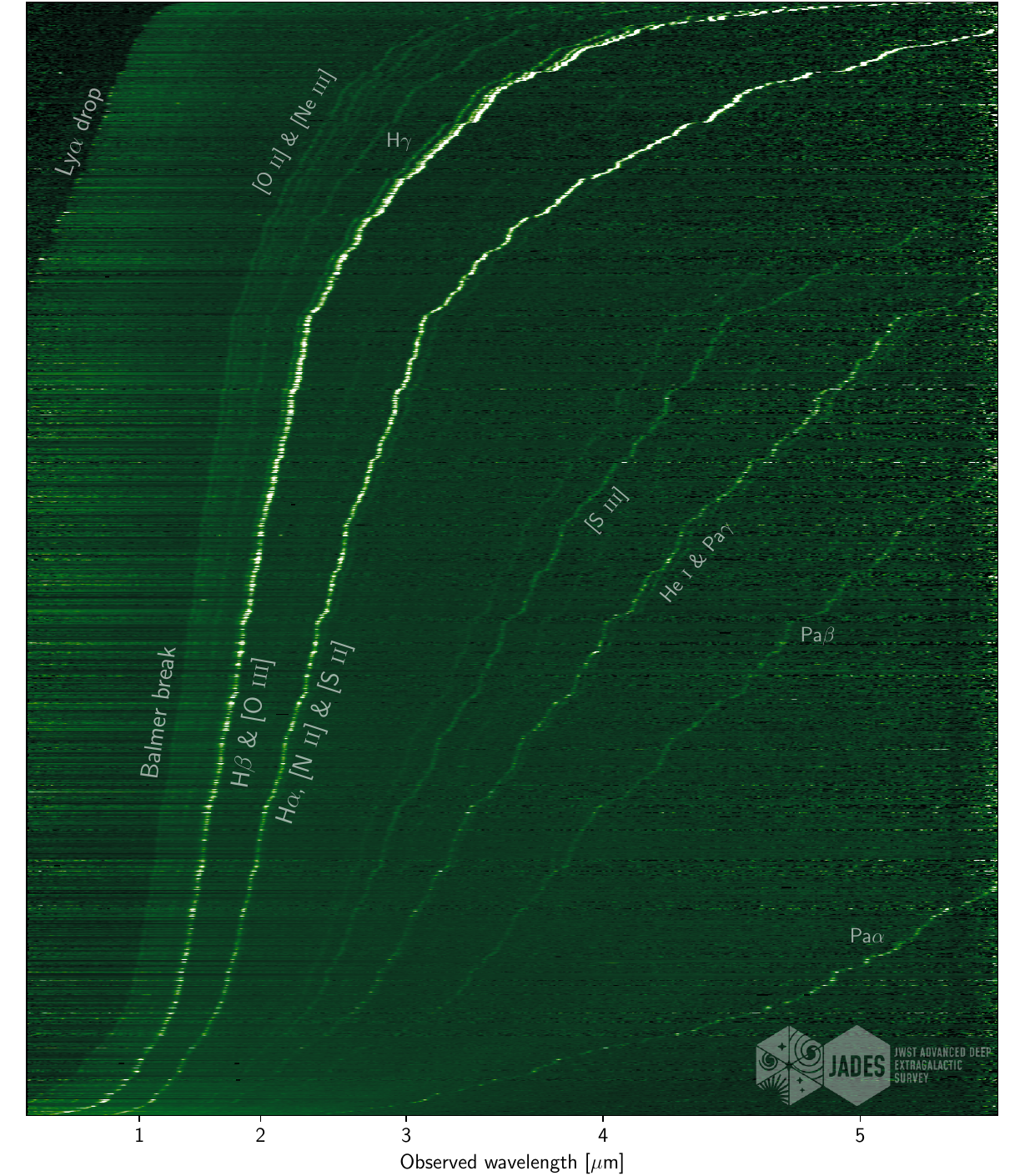} 
    \caption{A selection of high-quality emission-line spectra from this Data Release, sorted by redshift. Numerous emission line and continuum features are clearly visible, showcasing the richness of the dataset.}
    \label{fig:spectra_stack}
\end{figure*}

Throughout the survey, redshifts were determined from a combination of visual inspection and spectral fitting. Figure~\ref{fig:spectra_stack} shows a subset of spectra ordered by measured redshift. A full description of the process of redshift identification can be found in \citetalias{DR3} and \citetalias{DR4_paper2}.
As outlined in those references, spectroscopic redshifts were assigned a quality flag according to the following scheme:

(A) unambiguous redshift with at least one emission line detection in the medium-resolution grating;

(B) redshift with two or more emission lines detected in low-resolution Prism/Clear spectrum;

(C) clear redshift determined from the continuum, or from the continuum and a single Prism/Clear emission line;

(D) tentative redshift, determined from visual inspection; and

(E) no redshift.

We note that, as described in \citetalias{DR4_paper2}, this latest data release includes a re-calibration of the wavelength scale of all of our spectra, which slightly changes redshifts compared to \citetalias{DR1} and \citetalias{DR3}.
Throughout this paper we use these updated redshifts. Further details of this recalibration and its implications can be found in \citetalias{DR4_paper2}, but we note here that, for the purposes of assessing the quality of photometric redshifts, the changes are not significant, with corrections typically being $\Delta z \lesssim 0.01$. 

We also note that, while the $R\sim2700$ G395H spectra could, in principle, refine the redshifts even further beyond those derived from the $R\sim1000$ medium-resolution spectra, this is a level of precision beyond the needs of this paper. Given that not all tiers included G395H observations (see Table~\ref{tab:obssummary}), and that, where present, the spectral coverage is much narrower than that obtained by the combination of the three medium-resolution gratings, we do not consider the high-resolution gratings in the redshift determination in this paper.

\begin{figure}
    \centering
    \includegraphics[width=0.99\linewidth]{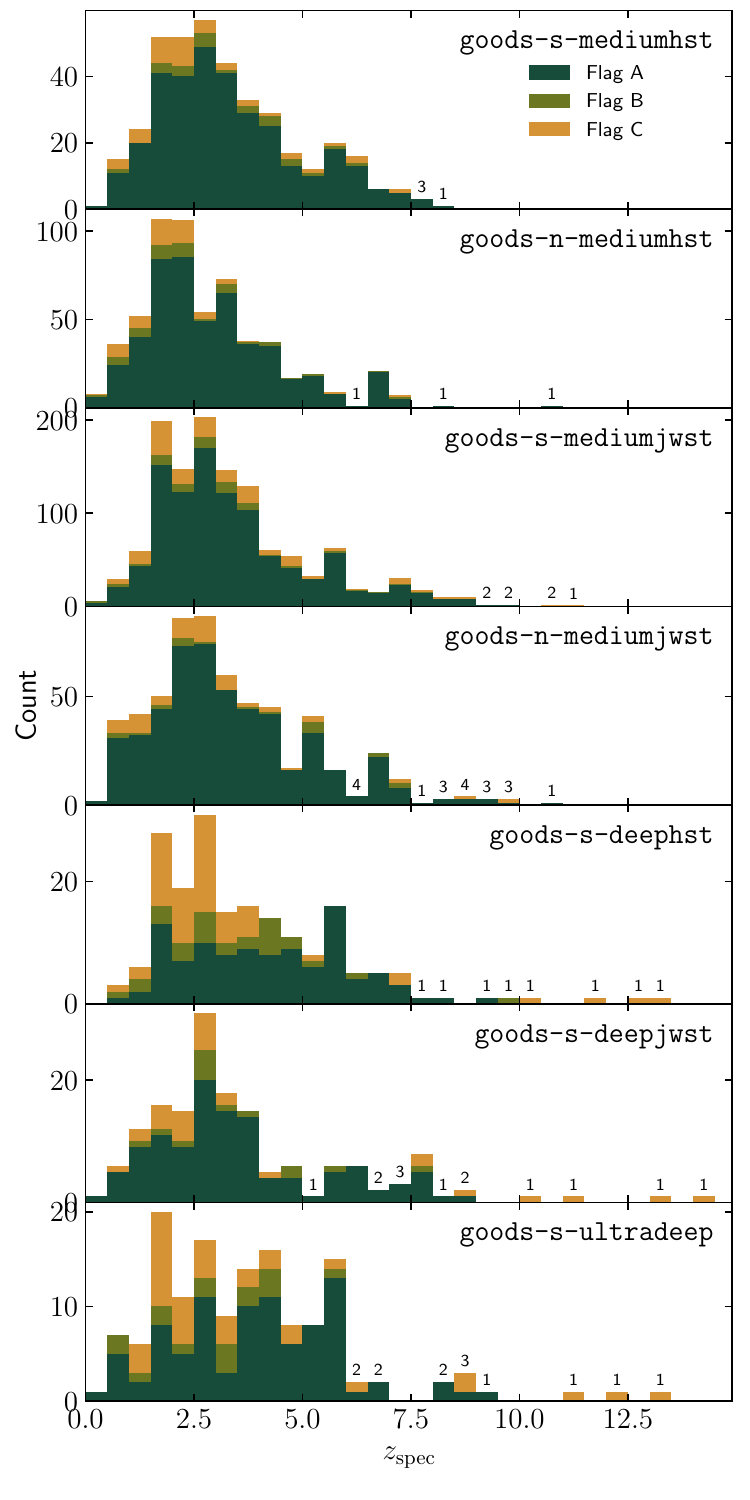}
    \caption{Histogram of all robust redshifts across each tier of the survey.}
    \label{fig:z_hist}
\end{figure}

Figure~\ref{fig:z_hist} shows a histogram of all robust (flag A, B, or C) redshifts in each tier of the survey.

\subsection{Quantifying success of target selection}
\label{sub:quantify_success}

In the following sub-sections, for each Tier of the JADES spectroscopic survey, we will consider the success rate for each priority class, i.e. what fraction of the galaxies targetted had robust spectroscopic redshifts (with quality flags A, B or C) in the anticipated range. As in our 
\citetalias{DR1} paper, we allow a buffer of $\Delta z=0.1$ beyond the target redshift range when counting a success. Galaxies with robust redshifts (quality flags A, B or C) outside this range are classed as interlopers. Galaxies without robust redshifts (quality flags D or E) are not counted as either a `success' or an `interloper'.

\subsubsection{Deep/\jwst}

\begin{figure*}
    \centering
    \includegraphics[width=0.65\linewidth]{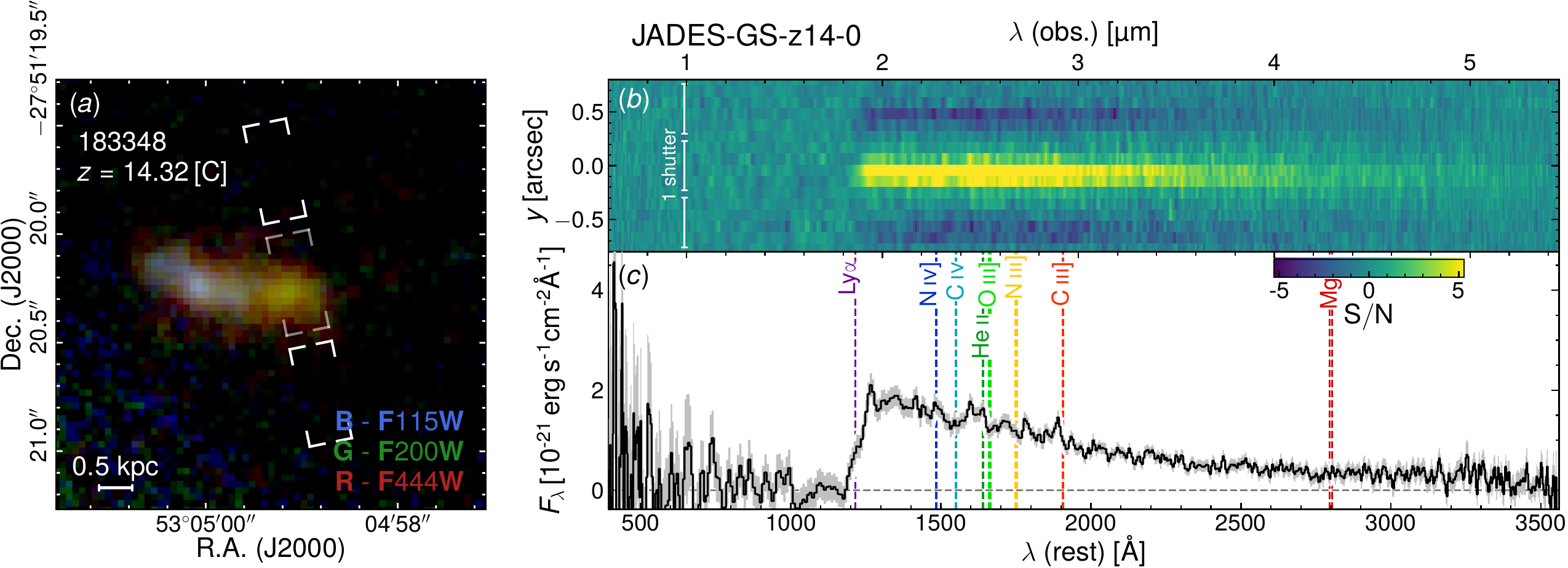}
    \includegraphics[width=0.65\linewidth]{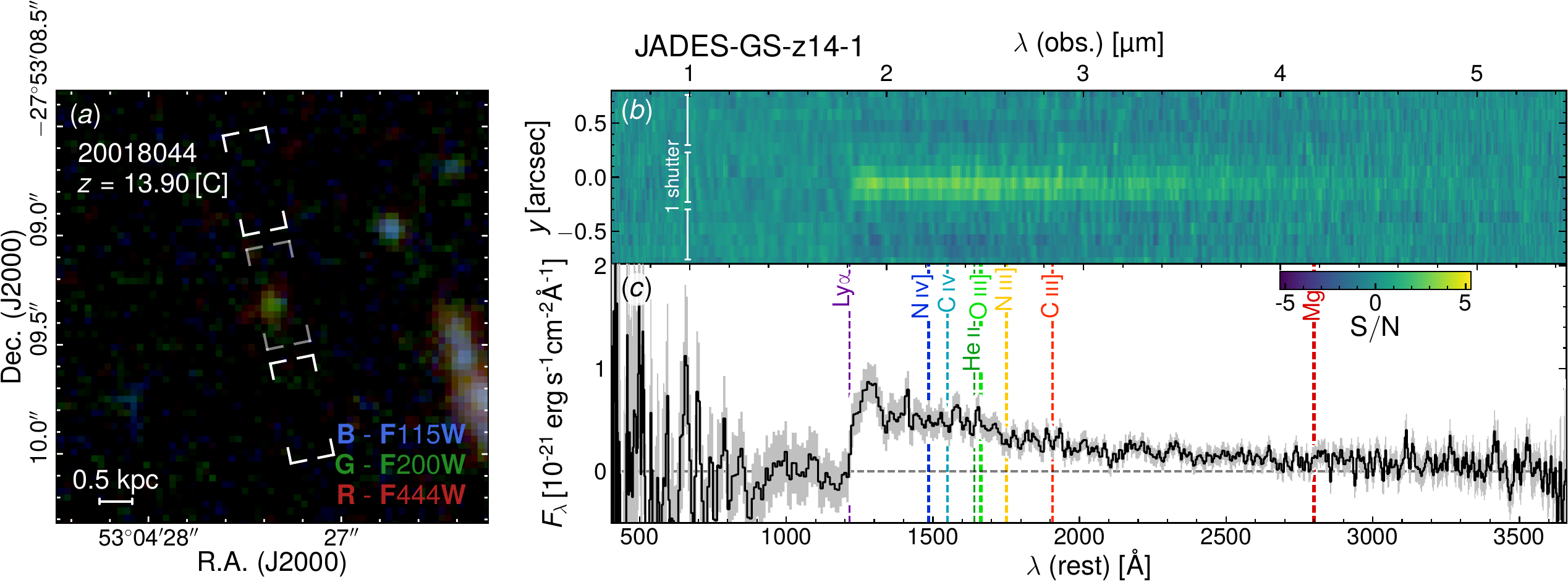}
    \includegraphics[width=0.65\linewidth]{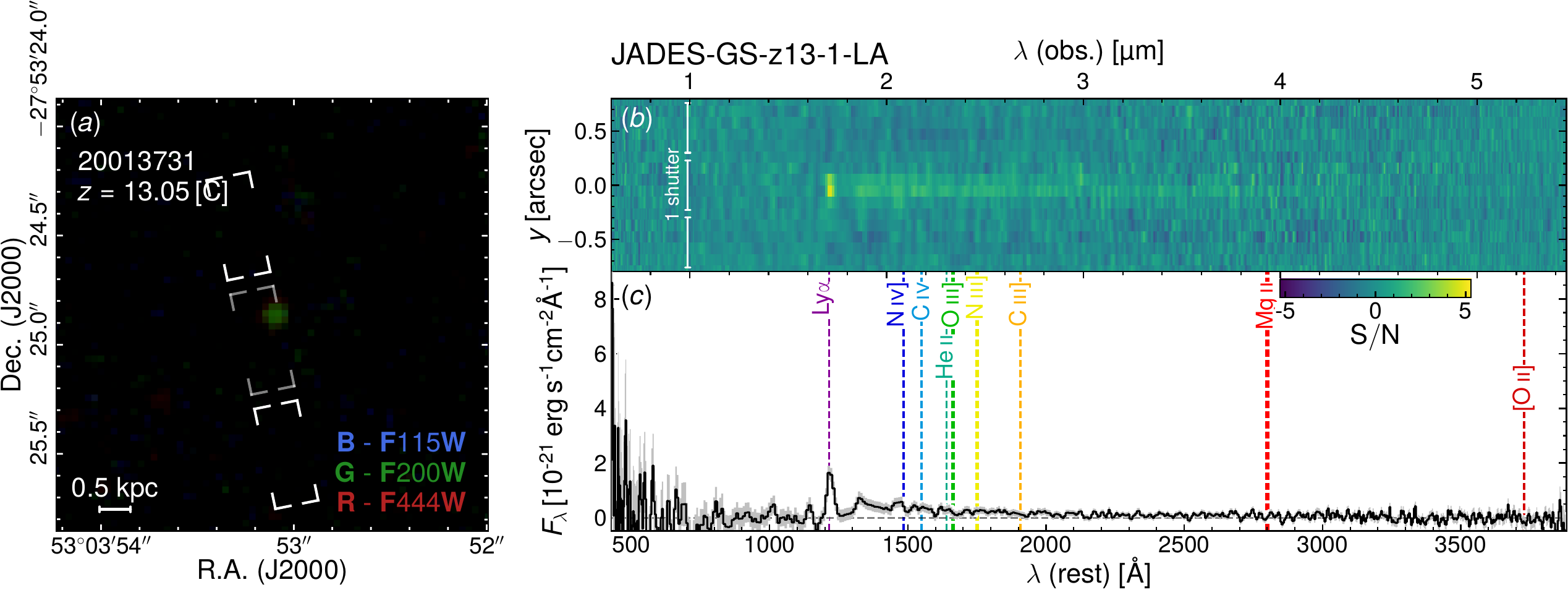}
    \includegraphics[width=0.65\linewidth]{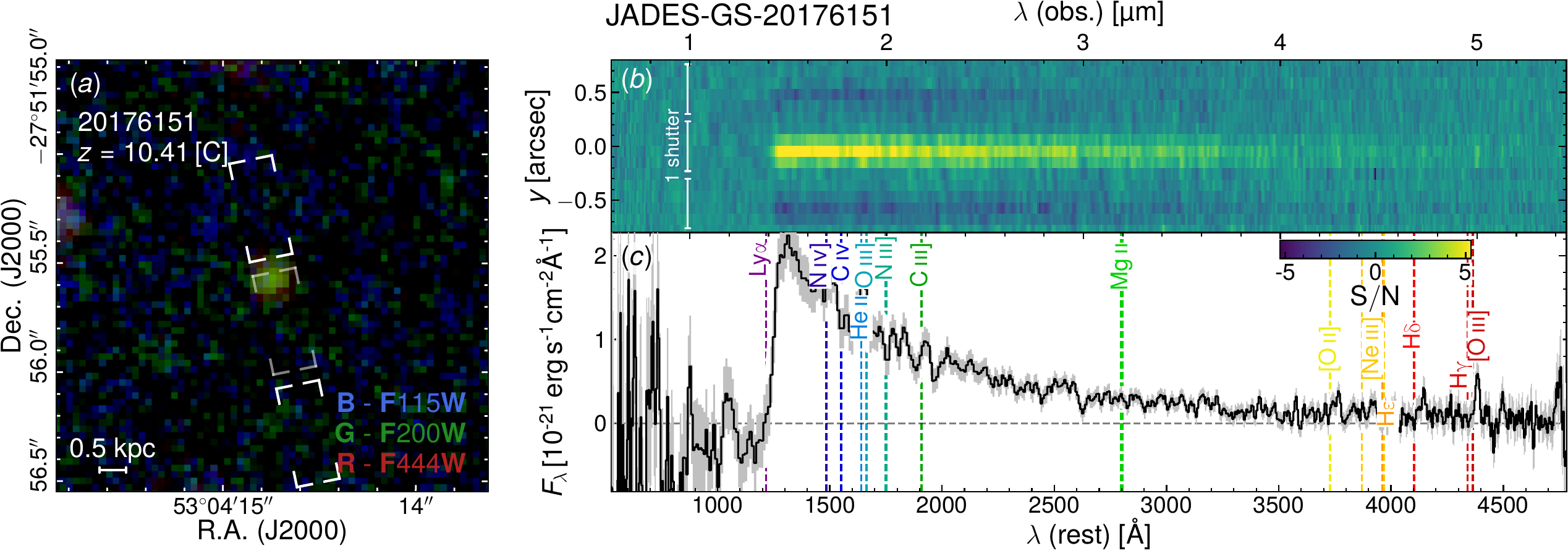}
    \includegraphics[width=0.65\linewidth]{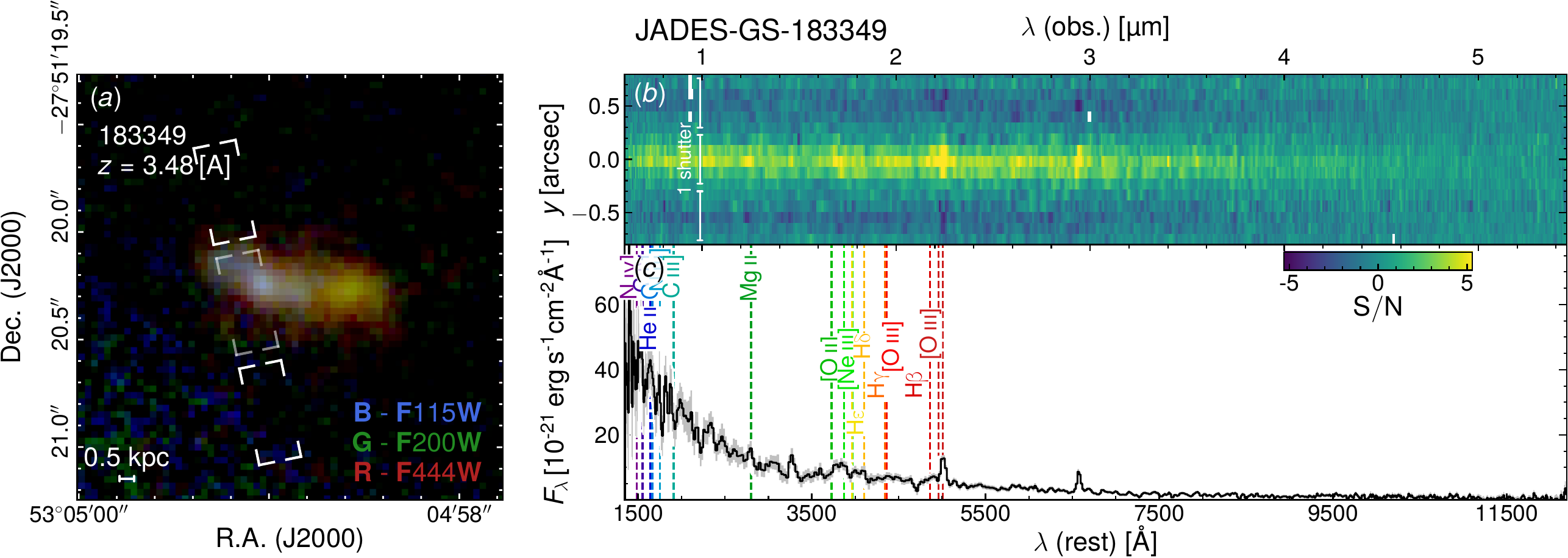}
    \caption{The spectra of objects with the highest priority (class 1) in \deepjwst that returned a spectroscopic redshift with flag `C' or above.  For each object plotted, panel (a) presents a false-color image from NIRCam, with the filter set displayed in the lower right corner. Panel (b) shows the two-dimensional signal-to-noise map from the NIRSpec/MSA prism data, where three shutters are marked. In this and subsequent 2D maps, regions of negative signal-to-noise arise from the background removal method, which relies on nodding and subtraction. Panel (c) displays the one-dimensional spectrum, extracted using a five-pixel boxcar method, as well as $1\sigma$ uncertainties shown as grey shading. We note that JADES-GS-183349 was a known low-redshift galaxy but was targetted in the highest priority class because of its proximity to JADES-GS-z14-0. }
    \label{fig:deepJWST_class1_FlagC}
\end{figure*}

\begin{figure}
    \centering
    \includegraphics[width=1\linewidth]{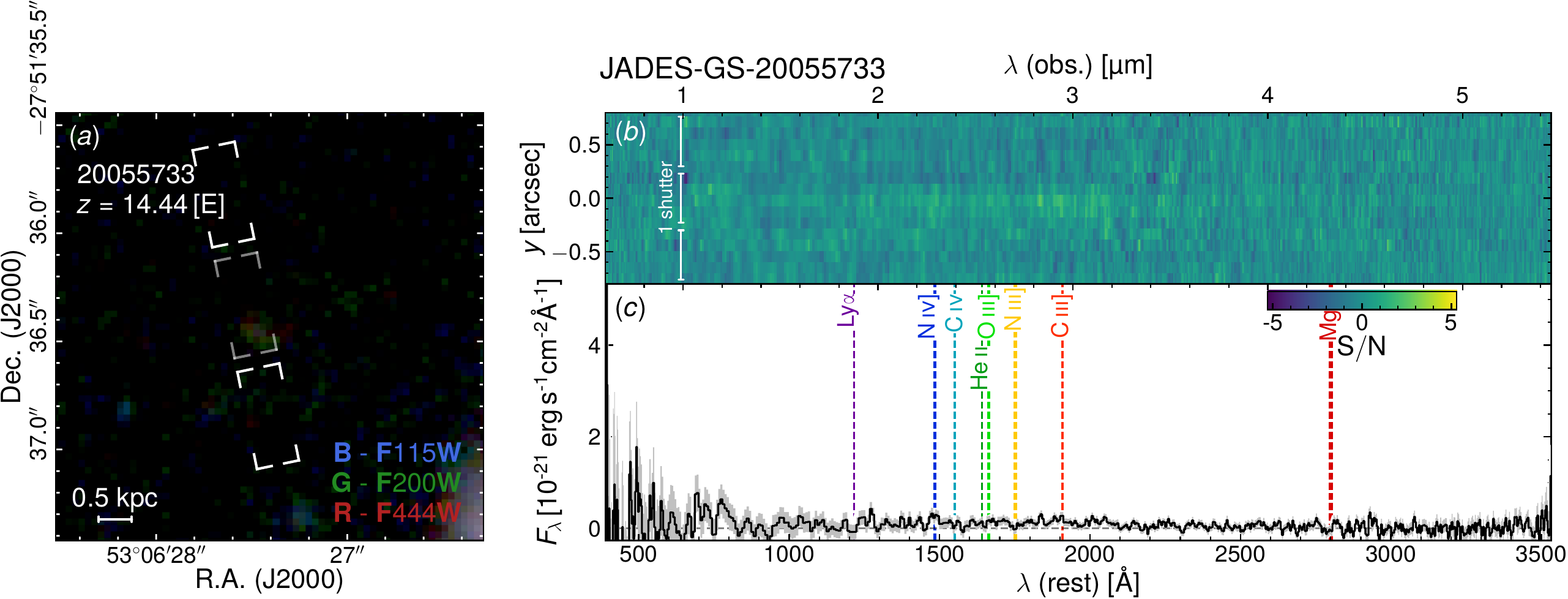}
    \includegraphics[width=1\linewidth]{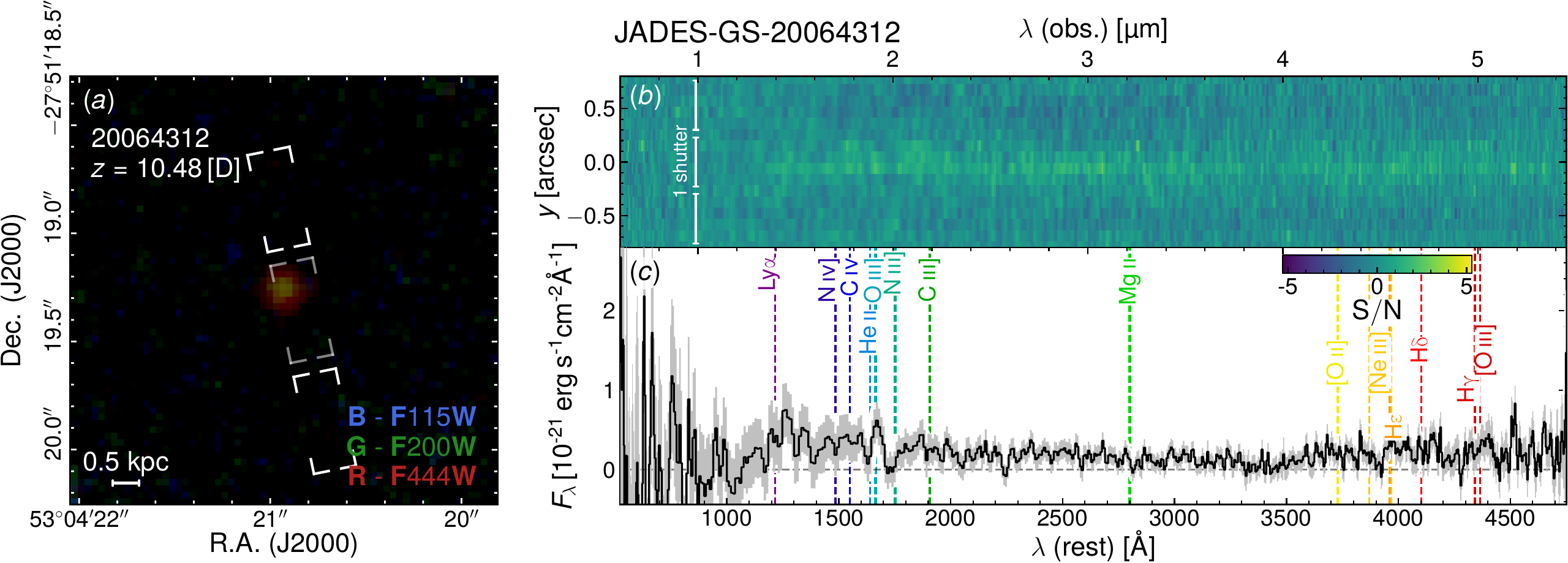}
    \includegraphics[width=1\linewidth]{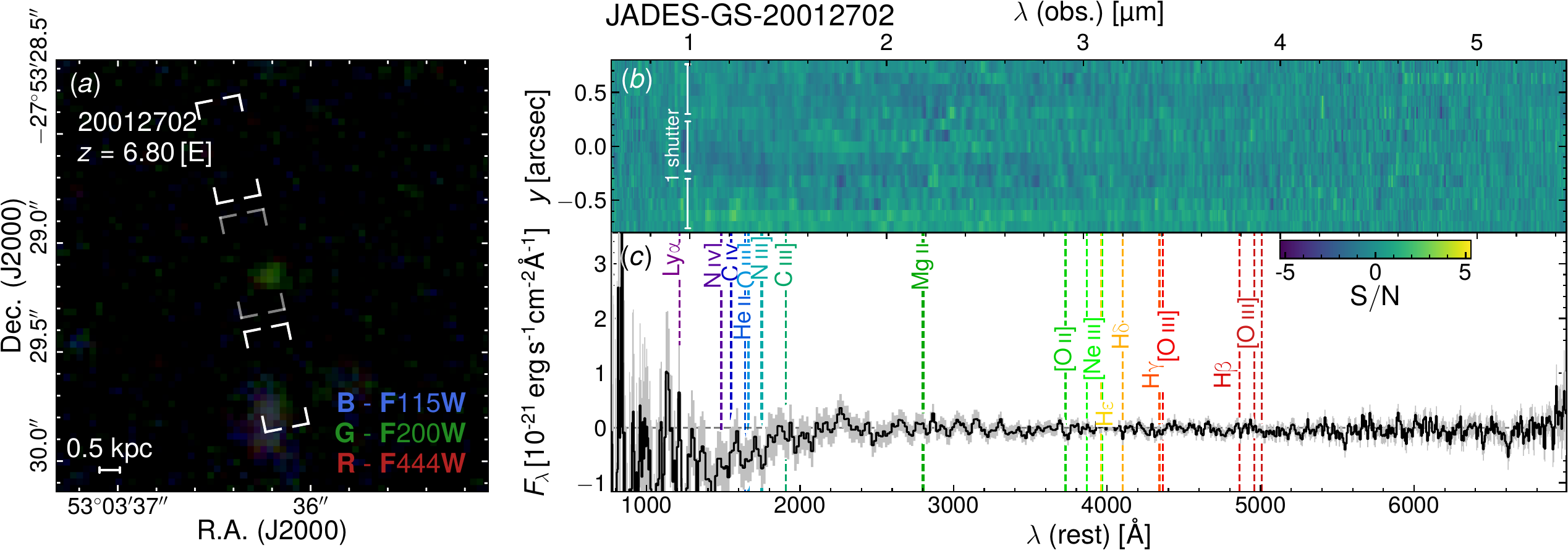}
    \caption{As for Fig~\ref{fig:deepJWST_class1_FlagC}, but for those spectra that did not return a firm redshift (visual inspection flags D-E) }
    \label{fig:deepJWST_class1_FlagD}
\end{figure}

For Deep-\jwst, we were ambitious in targeting relatively faint galaxy candidates beyond redshift 10 (Classes 1 \& 1.1). Ten objects were targeted, including the at-the-time highest redshift spectroscopically-confirmed galaxy at $z=14.18$ \citep{Carniani2024_GS_z14,Carniani2025,Schouws2025}, JADES-GS-z14-0. We also targeted a very nearby galaxy to GS-z14, which our broad-band photometry indicated was much lower redshift (a chance alignment on the sky) but we wished to check that GS-z14 was indeed high redshift and not a dusty region associated with the low-$z$ galaxy. We confirmed that the neighbouring galaxy (ID183349) is indeed at a lower redshift ($z=3.475$), but we exclude this from our ``success rate'' calculations. Of the 7 remaining high-redshift targets in Class 1, 4 (57 \%) were confirmed to have $z>10$ based on a clear Lyman break, but these typically lacked strong emission line detections (i.e.\ the spectra were flag C). We note that the $z=13$ galaxy ID 20013731 has a Lyman-$\alpha$ line \citep{Witstok2025_z13}, JADES-GS-z13-1-LA.  
Class 1.1 targeted 2 galaxies, of which one was a success, also based only on continuum features (flag C).
None of our targets in Classes 1 \& 1.1 were identified as clear low-redshift interlopers; of the remaining four that were not definitively confirmed as high-redshift (i.e. not flags A, B or C), two clearly showed flux and a break (flag D) which if due to the Lyman-break would be $z\approx 10$, but the redshift was not conclusive. Another two did not show sufficient flux to determine even a tentative redshift. 
The spectra of all Class 1 targets that returned robust redshifts are shown in Figure~\ref{fig:deepJWST_class1_FlagC}, and the three that did not yield robust redshift are shown in Figure~\ref{fig:deepJWST_class1_FlagD}.
One of these targets (ID 20012702) was positioned right on the edge of the shutter, and thus a significant amount of its flux would have been lost, although the other (ID 20055733) was reasonably well-centered.

In Class 2, 7 out of 13 (54\%) were successes with redshifts $z>7.9$, and one further galaxy (ID 20062446) was close to the target redshift interval at $z=7.65$ -- although formally the latter is considered an interloper in our statistics. There were two further interlopers (IDs 70378948 and 17909) at much lower redshift ($z\approx 2.1$) which showed a Balmer break in the spectrum whose wavelength would correspond to a Lyman-break at $z\approx 8$ (hence their selection as potential high-$z$ candidates), but in both cases the spectra showed flux short-ward of the break and weak [O{\sc iii}]~$\lambda$5007 and H$\alpha$ emission lines making the low-redshift identifications unambiguous. Two further galaxies (IDs 9442 \& 20015285) had continuum detections (Flag D) and evidence for a spectral break, but we could not definitively determine whether the break was a high-redshift Lyman-break or due to the Balmer/4000\AA\ break at much lower redshift. 

Class 3 goes even fainter than Class 2 but with the same target redshift range, and had just one target placed on the MSA. The spectrum showed some flux, but a robust redshift could not be determined. 

Class 3.1 aimed to target rare objects, such as high equivalent-width emitters, very blue galaxies, ALMA-detected galaxies, and AGN,  with $z>5.7$. In the end, the two galaxies targetted were both high equivalent-width emitters, selected based on photometric excess.
Both of these targets were confirmed as successes, with redshifts of
$z=7.96$ 
and 
$z=5.899$, 
respectively.

Our one target in Class 4 (which was for bright $z>5.7$ galaxies) ID 60318988 was a clear success with $z=6.05$ and a high S/N spectrum with good continuum and many emission lines. Classes 6.1 \& 6.2 target a similar redshift range to Class 4, but go fainter in the rest-UV. The success rates were 59\% for Class 6.1, dropping to 36\% for the fainter Class 6.2. The interloper rates were both $\approx 10$\%.

We note that Class 5 (bright galaxies at $z>2$) did not have any targets placed, and similarly Classes 7.1--7.4 (rare objects at $1.5<z<5.7$) 
also failed to have any targets placed on the Deep-\jwst MSA, due to low target density.

The remainder of Class 7 (7.5--7.9) comprises galaxies from $4.5<z<5.7$ to $1.5<z<2.5$ in $\Delta z=1$ slices, with a F444W magnitude cut. The highest-redshift slice only had three targets, with one success and one interloper. The other redshift slices had success rates above 70\%. These redshift slices all had interloper rates below 10\%, with the exception of Class 7.8 (brighter galaxies in the lowest redshift slice) which had an interloper rate of 18\% (two galaxies out of 11, both of which had $z=2.69$, only marginally above the target redshift range).

In Classes 8 \& 8.1, 54\% of targets did indeed prove to be above the target redshift threshold of $z=1.5$, and none had robust redshifts below this. Class 8.2 targetted galaxies below $z=1.5$, and 55\% were confirmed in this range, while 16\% returned redshifts above this. Classes 8.3 \& 9 had no redshift cut, and we were able to get redshifts for 8 out of 24 \& 1 out of 5 of galaxies in these classes respectively.

\subsubsection{Medium/\jwst}
\label{subsubsec:Medium/JWST}

Our top priority (Class 1) was aimed at targeting bright galaxies at the highest redshifts, but the exact criteria evolved as the survey progressed in time. 

In the first epochs prior to October 2023 (GSa, GN; see Table~\ref{tab:priorities_Medium_JWST}), this was delineated as $z>9$ and $m_{UV}<27.9$ and was selected based on photometric redshifts or Lyman-break dropout criteria (see Section~\ref{sec:target_selection}), with a visual inspection to retain just the most robust candidates (Class 1a), for which we had a had a 53 \% success rate. This class yielded a 33~\% interloper rate, however 4/5 of these interlopers had $z>8.3$. The less robust candidates from visual inspection were placed in priority Class 2a, of which only one was targeted and no redshift was obtained.
Slightly fainter targets with the same redshift cut were placed in Class 3a. The success rate for these faint targets was low (12~\%), although 2/3 interlopers came in with redshifts $z_{\rm spec}>7.3$.

In the later part of the survey (GSb; after September 2023), we primarily drew on catalogues from \citet{Hainline2024} and \citet{Endsley2024}, and lowered the redshift cut to $z>8$ with a  $m_{UV}<28$ cut to increase the target density (Class 1b). This was also a better match to the F115W-dropout colour-based technique. This proved successful, with a 76~\% success rate. The 21~\% interloper fraction was made up only of galaxies with redshifts $z>7.6$.
Fainter targets extending down to $m_{UV}<29.8$ from these catalogues were also included lower down our priority scheme as Class 3b, and this similarly returned a good success rates of 75~\% and featured no interlopers.

Targets that were not included in these catalogues, but returned $z_{\rm phot}>8$ formed Class 2c, and this had a much lower success rate of 20~\% (40~\% interloper fraction, although one of these two interlopers missed the redshift window only narrowly at $z=7.88$), reflecting that most of the robust high-redshift sources were selected in the \citet{Hainline2024} and \citet{Endsley2024} catalogues. 
Class 3c then extended this redshift cut to fainter magnitudes, and the success rate was also low for these targets (12~\%; with a 38~\% interloper fraction).

Some of the Medium/\jwst pointings in GSb were targeting the northern region of GOODS-S where we found there to be a low density of $z>8$ targets (see Figure~\ref{fig:GS_detail}). As a result we supplemented Classes 2 and 3 with F090W-dropout selected targets ($z\gtrsim6.5$) split by magnitude, with those having $m_{UV}<28$ going in Class 2d, and those with $28\leq m_{UV}<28.5$ going in Class 3d.
The former had a modest success rate of 47~\% (with a 20~\% interloper fraction), while the latter yielded only two (22~\%) successes, and one (11~\%) interloper, with the remainder not yielding redshifts.

In addition to our $z>8$ candidates, we included eight targets from a photometrically-identified overdensity at $z\approx7.3$ \citep{Saxena2023,Endsley2024,Witstok2024}, split between classes 1x and 2x, four of which had $z_{\rm spec}=7.26\pm0.01$, two had $z_{\rm spec}=7.48\pm0.01$, while the other two were at $z_{\rm spec}=7.00$ and $z_{\rm spec}=6.96$ respectively.

As expected, the bright high-redshift targets in Class 4 ($m_{AB}<26.5$ and $5.7<z<8$) had a high redshift confirmation rate of 88\%, with only 1 (6\%) interloper: a strong Balmer break at $z_{\rm spec}=1.27$.

Class 5 were very bright $z>2$ galaxies and also had a very high success rate of 85\% (17 out of 20) with just one interloper, a dusty galaxy at $z_{\rm spec}=1.67$. 

Class 6.1 is fainter than Class 4 but a similar redshift range, with a 60\% redshift success rate and a 10\% interloper rate.
Class 6.2 covers fainter magnitudes and it split into 6.2a, a $F444W<27$ mag cut, and 6.2b, a fainter version of the $m_{UV}$ cut used in 6.1 (magnitudes $28-28.5$). The fainter $m_{UV}$ cut in 6.2b yields slightly less favourable results than 6.1, with a success rate of 47\%, and an interloper fraction of 18\%. However the sample obtained via the F444W cut performs even worse with a 36\% interloper rate and only 21\% of targets returned a redshift in the desired range.

Class 7 were descending redshift slices from $4.5<z<5.7$ in steps of $\Delta z=1$ until $1.5<z<2.5$, with unusual galaxies (those obeying a UVJ cut for quiescence, or with X-ray emission) being placed first as Classes 7.1-7.4, followed by a simple magnitude cut of ${\rm F444W}<27$\,mag, to approximate a simple mass selection, in classes 7.4--7.9 (with the lowest redshift slice split into two magnitude bins in Classes 7.8 and 7.9).
We had a high success rate for the unusual galaxies at $1.5<z<4.5$ (Classes 7.2--7.4) of $\gtrsim 90$\%, although of the 3 targets in the highest redshift slice (class 7.1), 2 had redshifts below $z<4.5$.

For the more numerous `normal' galaxies selected from a F444W cut, we had a $\approx 70-80$\% success rate, with the exception of the faintest lowest redshift bin (Class 7.9 with 62\% success), and a low interloper fraction of $\approx 10$\%.

These high success rates compared with the corresponding Classes in Medium-HST (Classes 4 \& 5 which had success rates of $\approx 30-80$\%) shows the great strength in having the \jwst multi-wavelength photometry to get accurate photometric redshifts.
We note that the improvement of the photometry across the lifetime of the survey resulted in only a modest improvement in success rates. Table~\ref{tab:gsa_gsb_compare} shows the comparison between GSa and GSb success rates for Classes 7.5-7.8. Although the success rate grew significantly from GSa to GSb for Class 7.6 and Class 7.8, the change was very modest for Class 7.5 and Class 7.7. This highlights that, even with earlier reductions of \jwst/NIRCam imaging, the improvement of photometric redshift compared to pre-\jwst determinations was more significant than the improvements that came with later iterations of NIRCam data reduction.

Class 8.0 \& 8.1 pick up left-over galaxies with $z>1.5$, and, of the 651 galaxies, we confirm redshifts at $z>1.5$ for 39\%, whilst 1.5\% have lower spectroscopic redshifts.
Galaxies in class 8.2 have photometric redshifts $z<1.5$, 38\% of these 345 galaxies have spectroscopic redshifts confirming this, with another 12\% having higher spectroscopic redshifts.
Class 8.3 has no redshift cut (and F444W brighter than 29\,mag) and we get a redshift for 21\% of 280 sources.

\subsubsection{Ultra Deep}

For the Ultra Deep spectroscopy, taken in programme GO-3215 (PI: D.\ Eisenstein), the strategy was different, with 168ks on source, and pushing far deeper in one of the $R\sim1000$ gratings (G395M) rather than obtaining the full wavelength coverage in all three gratings.  We therefore extended our spectroscopy to encompass fainter targets down to $m_{AB}=30$\,mag in an MSA pointing which repeated the Deep-HST field \citepalias{DR1}, but with some changes in targets. In class 1, we repeated three of the $z>11$ galaxies we had initially obtained spectra for in the Deep-HST tier, which were first reported in \citet{Curtis-Lake2023}. The new Ultra-Deep spectrum of GS-z12 has been published in \citet{DEugenio2024_Carbon}, and those of GS-z11 and GS-z13 have appeared in \citet{Hainline2024b}. A fourth galaxy which was placed in the highest priority Class 1 was the $z=9.43$ galaxy identified in our Deep-HST spectroscopy \citep{DR1,Cameron2023_line_ratios}. The Ultra-Deep spectrum of this galaxy, GS-z9-0, has been presented in \citet{Curti2024_GS_z9}. 
Since the redshifts of these galaxies had already been spectroscopically confirmed, we do not discuss the success rates here.
Classes 1.2 \& 2.1 contained other candidate $z>10$ galaxies with lower priorities, but we were unable to place any of these on our MSA design.

Class 2.3 included targets at $8<z<10$, and 5 of 6 (83\%) were spectroscopically confirmed in this range, with the other target inconclusive. Class 2.4 covered the same redshifts but had a less strict cut on the quality of the photometric redshift, but we were unable to get robust redshifts for either of the two placed targets.

Classes 3.1 \& 3.2 selected for ``rare objects'', including candidate AGN, quiescent galaxies, galaxies with ALMA detections, extremely blue galaxies and more. These spanned a wide range of expected redshifts, but, generally speaking, brighter and higher-redshift rare objects were put into to the higher of the two classes.

Five objects from Class 3.1 were observed; 60~\% of this class yielded redshifts that were in line with expectation, while the remainder did not yield a redshift. Two bright quiescent galaxies at $z\sim2.8$, both of which came in at approximately this redshift, thus, we deem these successes. Two extremely blue candidates from \citet{Topping2024} were observed, one came in with $z_{\rm spec}=5.961$, but the other did not yield a redshift. The final source was a candidate AGN that did not yield a redshift.
Note that we do not consider the true nature of the object when reporting these success statistics (i.e., whether the object was indeed quiescent).

A further four objects were observed in Class 3.2. Three of these yielded redshifts in line with expectation: a candidate AGN at $z_{\rm spec}= 5.560$, and ALMA-detected galaxy from \citet{Decarli2016} at $z_{\rm spec}= 2.843$, and a passive galaxy candidate at $z_{\rm spec}= 1.884$ from \citet{Cassata2013}. The other object in Class 3.2 was a candidate extreme line-emitter that did not yield a redshift.

All three of the targets selected in Class 4.1 came in with redshifts in the desired range of $5.7<z<8$.
For Class 4.2, the success rate was lower, but we note that while the majority of targets were classed as `interlopers', their redshifts were actually at $5<z<5.6$, so only narrowly miss the formal cut-off for `success'. This Class 4.2 was derived from F775W dropout catalogues from \citet{Endsley2024}, and the photometric redshift histogram (their Figure 1) extends down to $z\approx 5.0$. If we would instead consider the range $5<z<7$, this class would have a 88\% success rate. 

Lower down our priority selection, Class 6.1 encompassed leftover targets with $5.7<z_{\rm phot}<8$ passing the same magnitude cut as Class 4.2, which were {\em not} selected in the \citet{Endsley2024} sample. Perhaps unsurprisingly, this category of residual targets had a success rate of 0 \%, suggesting the \citet{Endsley2024} selection had a high completeness.

Classes 5.1 \& 5.2 had a reasonable ($\sim60$~\%) success rate in selecting $4<z<5.7$ targets with interloper fractions below 25~\%. 
Interestingly, three out of the five interlopers from Class 5.2 were actually above the target redshift range with redshifts of $z_{\rm spec}=5.9, 6.0, {\rm and~} 6.9$.
Class 6.2, which extended this redshift range down to $29<m_{AB}<30$, had a far more modest success rate of 24~\%, with an almost 40~\% interloper fraction.

For the lower redshift slices targeting $25<m_{AB}<28$ galaxies (Class 7.1 \& 7.3), the success rate was extremely high ($>85$~\%), but slightly fainter classes selecting $28<m_{AB}<29$ (Class 7.2 \& 7.4) had more modest success rates of 67\% and 40\% respectively. Interestingly, none of Class 7.1-7.4 yielded any clear interlopers, although many objects did not result in any robust redshift.

Classes 8.1-8.3 placed $z<1.5$ candidates as filler targets, divided into three different magnitude bins. The brightest of these (8.1; $25<m_{AB}<28$) yielded redshifts for just over half the placed objects, with 48~\% confirmed to have $z<1.5$, while just under 10~\% turned out to be at higher redshifts. A total of 12 objects were placed in 8.2 and 8.3 ($28<m_{AB}<29$ and $29<m_{AB}<30$, respectively), with only one being confirmed with $z<1.5$ and six turning out to be $z>1.5$.
Nine filler objects from our HST-based catalog were allocated in Class 9 of which one yielded a redshift ($z=0.334$).

\subsubsection{Medium/HST}

\begin{figure}
    \centering
    \includegraphics[width=0.99\columnwidth]{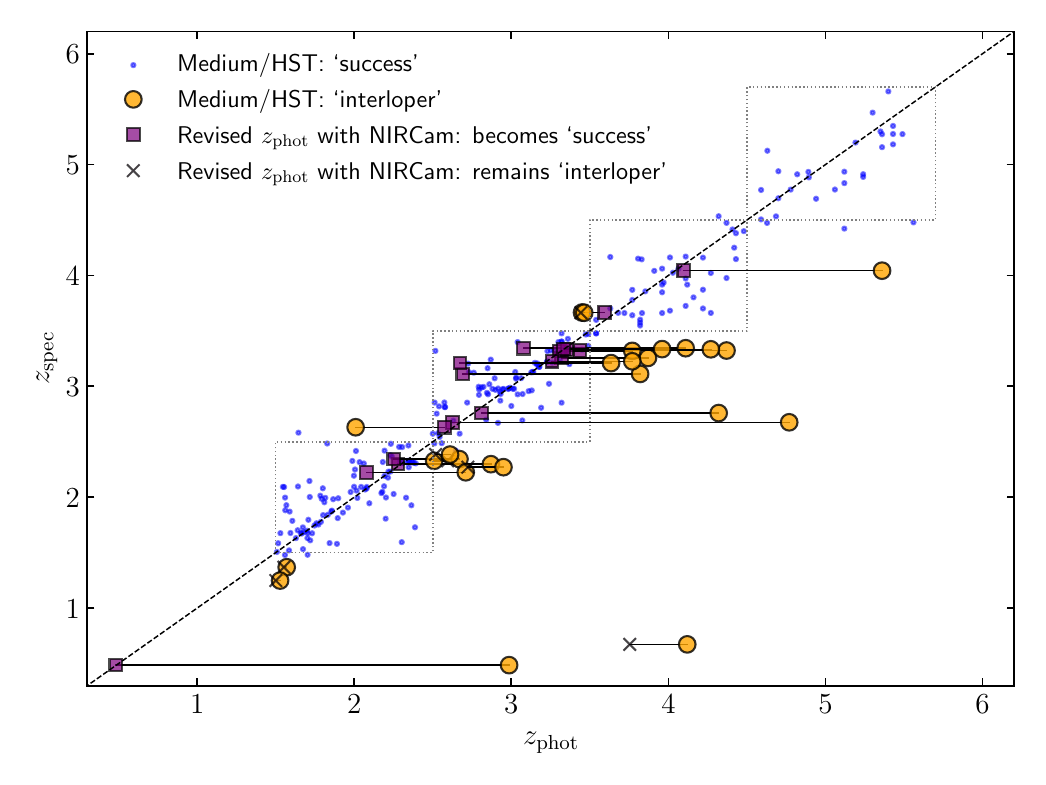}
    \caption{Improvement of the interloper $z_{\rm phot}$ after addition of NIRCam for HST-selected targets from Medium/HST}
    \label{fig:medHST_withNIRCam}
\end{figure}

The Medium/HST class scheme is discussed in \citetalias{DR3}, and summarized here in Table~\ref{tab:priorities_Medium_HST}.
Since all Medium/HST target selection was performed before \jwst/NIRCam pre-imaging was available and the spectroscopic exposures were reasonably shallow, we did not have a specific $z>8$ class, and our highest class was instead only aimed at selecting galaxies with $z>5.7$. 
Our most robust, brightest candidates in this redshift range (Class 1) have a high confirmation rate of 90\% (Table~\ref{tab:priorities_Medium_HST}), with a slightly lower 61\% success rate in Class 2 (which were either fainter or less robust on visual inspection). Class 1 had a low interloper rate of 6\%, while Class 2 had a slightly higher 11\%. 
The high success rate is probably related to the reliability of the Lyman break as a redshift indicator in the sensitive HST-ACS filters, with $z>5.7$ corresponding to $i'$-band dropouts. 

Classes 3.0 \& 3.5 of Medium/HST consist of rare targets at $1.5<z<5.7$, such as AGN and quiescent galaxies (Class 3.0) or bright sources (F160W$<23.5$\,mag; Class 3.5). The success rate was very high, as expected, with 85\% of 20 in Class 3.0 (with one interloper) and 100\% of 37 in Class 3.5.

In each of the Classes 4, 5 and 6 we consider slices in redshift, descending from $4.5<z<5.7$ down to $1.5<z<2.5$ in $\Delta z=1$ increments. Class 4 considers brighter galaxies with \hst F160W$<25.5$\,mag, and Class 5 goes fainter to $27$\,mag. Class 6 comprises leftover objects which did not make the F160W flux cuts but for which the SEDs suggested that there may be sufficient star formation rate (SFR) for line emission detection.

For the brighter galaxies in Class 4, both the success rate and interloper fraction improve as we go from the highest redshift slice ($4.5<z<5.7$) to the lowest ($1.5<z<2.5$). Sub-class 4.1 has a particularly large interloper fraction (33\%). We note that, given the selection was \hst-only, the reddest broadband filter (F160W) only covers until $\lambda_{\text rest} \sim 3000$\ \AA{} for this redshift interval. Given that this class specifically selected for the brightest candidates in F160W, it is perhaps not overly surprising that a reasonable number of those with photometric redshifts in the highest redshift bin  turned out to be lower-redshift interlopers (as they would have been extremely luminous at the target high redshift). 
For the slightly fainter sample in Class 5, the success rate  of $\approx 55-70$\% and interloper fraction of $\approx 9$\% are quite stable across the redshift slices.

For all interlopers identified across Classes 4 \& 5 that subsequently received NIRCam imaging, we found that their new photometric redshifts were generally much better aligned with their spectroscopic redshifts. As shown in Figure~\ref{fig:medHST_withNIRCam}, we find that the majority of these interlopers would have been placed in the `correct' redshift bin if the selection was repeated with this NIRCam photometry in hand. This highlights the value of broad wavelength coverage in SED fitting.
We also note that objects selected in this class were not subject to the same level of visual inspection as higher classes.

In Class 6 (faint galaxies not meeting the F160W cut) the success rate was lower (up to 55\% for $2.5<z<4.5$ but dropping to $\approx 25$\% in the lowest and highest redshift slices) with a moderate interloper fraction (22 \%) in the highest redshift slice ($4.5<z<5.7$) and low interloper fractions of $\approx 9$\% for the other redshift slices. We note that many of Class 6 did not have robust redshifts, not unexpectedly given their faint magnitudes, which precluded continuum detection and also rendered the SED fitting from the broad-band photometry uncertain. However, having this Class did indeed enable us to capture some galaxies with strong line emission and weak continuum which would have escaped selection in Classes 4 \& 5 over this redshift range.


We were able to confirm redshifts for 39\% of the leftover $z>1.5$ candidates placed in Class 7. Of the bright $z<1.5$ candidates placed in Class 7.5, 34\% were confirmed to have redshifts in this range, although 11\% were in fact found to have $z>1.5$, while the fainter targets in Class 7.6 had a very low rate of identifiable redshifts. Class 8 contained any remaining objects which would not saturate the detector and could be placed on the MSA, and of those 102 objects we identified robust redshifts for about half.

\subsection{Summary of success rates}

Overall, our redshift success rates were very high. 
For our F444W-selected objects in the range $1.5<z<5.7$, success rates were $\sim 80$\% with low interloper fractions of $\sim 10$\%, with some exceptions. This shows the photometric redshifts used were typically good, and adding NIRCam data clearly improved success rates over the \hst-only target selection.

For galaxies around the epoch of reionisation ($5.7<z<8$), we were generally successful in obtaining redshifts in the correct range for $\gtrsim60$~\% of targets when making selection cuts of $m_{\rm UV}\lesssim28$. However, these success rates begin to drop off as we pushed to fainter magnitudes. These high success rates reflect the robustness of using the Ly-$\alpha$ break in determining photometric redshift.

For the very high redshift candidates ($z\gtrsim8 -10$) in the top priority classes, we successfully confirmed $\sim60$~\% of these, although around $\sim20$~\% of our candidates -- typically the fainter sources that we placed -- did not yield robust redshifts.

\subsection{UV Luminosity Function}
\label{sec:uvlf}

\begin{figure}
    \centering
    \includegraphics[width=0.99\linewidth]{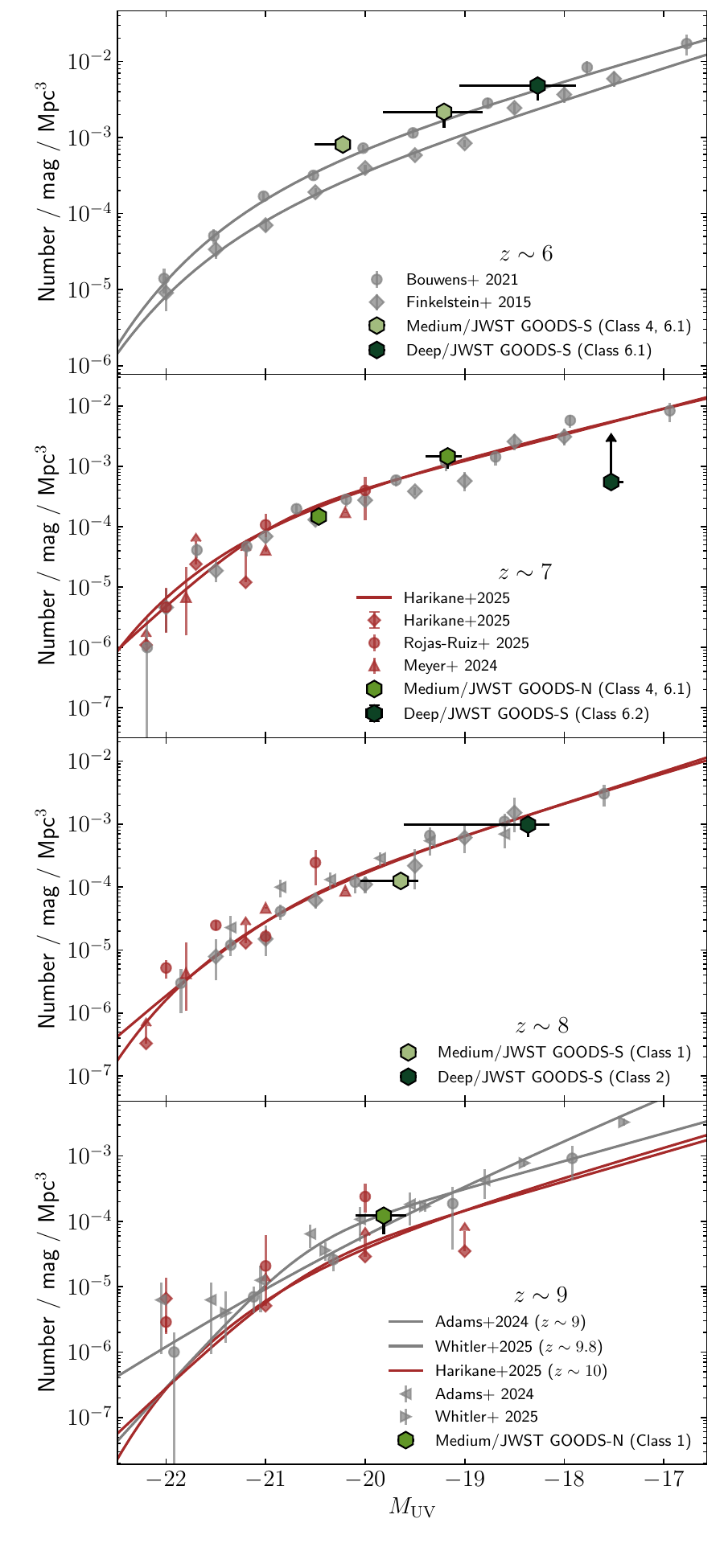}
    \caption{Spectroscopic UV luminosity function constraints for $z\sim6-9$ from the target number densities and spectroscopic redshift success rates described in this paper. Our values are in good agreement with luminosity functions from the literature based on spectroscopic redshifts (red points and lines; \citealt{Meyer2024, Harikane2025, RojasRuiz2025}) and photometric redshifts (grey points and lines; \citealt{Finkelstein2015, Bouwens2021, Adams2024, Whitler2025}).
             }
    \label{fig:uvlf}
\end{figure}

\begin{table*}
\centering
\caption{Constraints on the UV luminosity function derived based on number counts and redshift success rates from Medium/\jwst and Deep/\jwst.}
\begin{tabular}{ c c c c c c c c }
\hline
Tier & Field & Priority Class & $N_{\rm obj}$ & $z_{\rm median}$ & $F_{\rm comp,1}$~$^\dagger$ 
& $M_{\rm UV}$ & $\phi$ \\
 &  & & & & 
 ($F_{\rm comp,2}$) & & $10^{-3}$ mag$^{-1}$ Mpc$^{-1}$ \\
\hline
$z\sim6$ & & & \\
Medium/\jwst & GOODS-S & 4 & 3 & $5.920$ & $0.145$ ($0.155$) 
& $-20.22_{-0.29}^{+0.02}$ & $0.808_{-0.135}^{+0.084}$\\
Medium/\jwst & GOODS-S & 6.1 & 19 & $6.000$ & $0.105$ ($0.159$) 
& $-19.21_{-0.61}^{+0.39}$ & $2.17_{-0.838}^{+0.09}$\\
Deep/\jwst & GOODS-S & 6.1 & 8 & $6.308$ & $0.082$ ($0.122$) 
& $-18.27_{-0.79}^{+0.39}$ & $4.803_{-1.759}^{+0.176}$\\
\\
\hline
$z\sim7$ & & & \\
Medium/\jwst & GOODS-N & 4 & 2 & $6.842$ & $0.177$ ($0.189$) 
& $-20.47_{-0.03}^{+0.03}$ & $0.147_{-0.028}^{+0.019}$\\
Medium/\jwst & GOODS-N & 6.1 & 20 & $6.607$ & $0.120$ ($0.183$) 
&  $-19.17_{-0.22}^{+0.14}$ & $1.461_{-0.544}^{+0.04}$\\
Deep/\jwst & GOODS-S & 6.2 & 4 & $7.091$ & $0.064$ ($0.161$) 
& $-17.53_{-0.05}^{+0.12}$ & $>0.55$\\
\\
\hline
$z\sim8$ & & &\\
Medium/\jwst & GOODS-S & 1 & 20 & $8.423$ & $0.791$ ($0.822$) 
& $-19.64_{-0.41}^{+0.18}$ & $0.125_{-0.029}^{+0.025}$\\
Deep/\jwst & GOODS-S & 2 & 7 & $7.954$ & $0.140$ ($0.199$) 
& $-18.36_{-1.25}^{+0.21}$ & $0.976_{-0.35}^{+0.059}$\\
\\
\hline
$z\sim9$ & & &\\
Medium/\jwst & GOODS-N & 1 & 7 & $9.633$ & $0.804$ ($0.897$) 
& $-19.81_{-0.28}^{+0.23}$ & $0.122_{-0.058}^{+0.045}$\\
\\
\hline
\end{tabular}
\\
$^\dagger$ See Equations~\ref{eq:fcomp_case1}~\&~\ref{eq:fcomp_case2} for the definition of the two bounding cases of completeness fraction $F_{\rm comp,1}$ and $F_{\rm comp,2}$.
\label{tab:LF_values}
\end{table*}

Using the number density of photometrically selected targets from our parent sample and the spectroscopic redshift success rates outlined earlier in this section, we are able to construct a UV luminosity function for our sample of galaxies.
We focus here on Medium-\jwst and Deep-\jwst as these tiers are \jwst-selected and have the most uniform rest-UV based selection. From Deep-\jwst we consider Class 2 ($z>8$) and Classes 6.1 \& 6.2 ($5.7<z<8$). For Medium-\jwst, we consider Classes 4 \& 6.1 ($5.7<z<8$) and Class 1 (which in GOODS-North is $z>9$, and for GOODS-South is $z>8$).
We omit the early observations from GOODS-South (`GSa' in Table~\ref{tab:priorities_Medium_JWST}) because they had a significantly different parent catalogue from an earlier reduction of the NIRCam imaging.

Although our MSA-based survey is incomplete, we keep track of all galaxies in the ``parent sample'' (being all those within the MSA footprint),  
including those that were not ultimately observed (numbers in parenthesis in Tables~\ref{tab:priorities_Deep_JWST}-\ref{tab:priorities_Deep_HST}). Hence we are able to correct for the incompleteness in our survey.

We use the $1/V_{\mathrm max}$ technique \cite{Schmidt1968} to calculate the number density of galaxies in a particular priority class, only counting the galaxies with a robust spectroscopic redshift (flags A, B \& C) within the target redshift range for that class (corresponding to the `Success' column in Tables~\ref{tab:priorities_Deep_JWST}-\ref{tab:priorities_Deep_HST}).
In the calculation of $V_{\mathrm max}$, we consider the redshift range over which the galaxy would have been selected -- bounded by the lower redshift
cut of the priority class, and with 
$z_{\rm max}$ 
set either by the top end of the redshift range, or the redshift at which the galaxy would become fainter than the limiting apparent magnitude of the priority class selection (where we use a spectral slope $\beta= -2$, i.e.\ flat in $f_{\nu}$, and account for luminosity distance).
The area used in the volume calculation is that provided for each tier in Table~\ref{tab:areas}, which corresponds to the area union of all MSA footprints across the tier in question, discounting any area without NIRCam coverage.

After obtaining a number density $\phi$ by summing the $1/V_{\mathrm max}$, 
we correct for incompleteness by applying a correction factor $F_{\rm comp}$ based on the redshift success rate and fraction of the parent sample that was observed,

\begin{equation}
    \phi  = \frac{1}{F_{\rm comp}} \sum{}{}{\frac{1}{V_{\rm max}}}.
\end{equation}

As described in Section~\ref{sub:quantify_success}, targets which were not a redshift `success' could either be an `interloper' (robust redshift, outside of range), or in some cases returned no reliable redshift. 
In calculating $F_{\rm comp}$, we want to exclude interlopers from our sample, and we assume that the parent sample has the same interloper fraction as our observed sample. To handle galaxies which did not return a redshift, we consider two bounding cases. 
In one case, we assume the observed galaxies without redshifts are all interlopers. Adding them to the interloper fraction, we calculated $F_{\rm comp}$ as,  
\begin{equation}
\label{eq:fcomp_case1}
    F_{\rm comp, 1} = \frac{N_{\rm success}}{N_{\rm parent}\times N_{\rm success}/N_{\rm obs}},
\end{equation}

where $N_{\rm success}$ is the number of redshift `successes' among the observed sample, $N_{\rm parent}$ is total the number of candidates within the survey area, and $N_{\rm obs}$ is the number of targets for which we obtained spectra.\footnote{$N_{\rm obs}$ and $N_{\rm parent}$ are given for each class in the `allocated' and `possible' targets column of Tables~\ref{tab:priorities_Medium_HST}-\ref{tab:priorities_Deep_HST}.  In addition, the probability that a source is assigned a shutter is calculated from these values and stored on a source-by-source basis in the catalogues within the data release. The `Success' column reports $N_{\rm success}/N_{\rm obs}$.}

In the other bounding case we assume that the galaxies which did not return redshifts all fall within the target redshift range. 
In this case, $F_{\rm comp}$ is calculated as
\begin{equation}
\label{eq:fcomp_case2}
    F_{\rm comp, 2} = \frac{N_{\rm success}}{N_{\rm parent}\times(1-N_{\rm interloper}/N_{\rm obs})}
\end{equation}

where $N_{\rm interloper}$ is the number of redshift `interlopers' among the observed sample.\footnote{The `Interloper' column in Tables~\ref{tab:priorities_Medium_HST}-\ref{tab:priorities_Deep_HST} reports $N_{\rm interloper}/N_{\rm obs}$.} 
We adopt the second case (Equation~\ref{eq:fcomp_case2}) as our fiducial value, which returns a higher number density. 
We calculate a formal uncertainty on $\phi$ by summing $1/V_{\rm max}$ in quadrature, and we further add the difference between the two bounding cases into the lower error bar on our $\phi$ values. 
Table~\ref{tab:LF_values} reports the derived values for each tier and class that we consider, including the median and 16--84 percentile range on $M_{\rm UV}$ for the galaxies counted in each Class, which is what we use to plot our resulting values in Figure~\ref{fig:uvlf}.

Note that the different tiers span different ranges of $M_{UV}$ with some overlap, so the points plotted from JADES do not reflect different absolute magnitude bins, but do include independent samples of galaxies. We have not explicitly included cosmic variance in the error bars, but we are able to compare GOODS-North to GOODS-South to mitigate against this.
We note that our faintest bin at $z\sim 7$ (Deep-\jwst Class 6.2) shows a decline in density from the brighter bins. This is due to the magnitude cut being faint enough that the parent sample from NIRCam is likely not complete. Thus, we treat this value as a lower limit.

We compare with luminosity functions in the literature, split by redshift range. 
Grey points in Figure~\ref{fig:uvlf} show values derived from purely photometric samples \citep{Finkelstein2015, Bouwens2021, Adams2024, Whitler2025}. Red points show samples that use spectroscopic redshift inputs as well \citep{Meyer2024, Harikane2025, RojasRuiz2025}.

Overall, our values are in good agreement with both photometric and spectroscopic determinations. 
This suggests that previous photometric-based determinations of the UV luminosity function have not been strongly impacted by redshift interlopers. This reflects the robustness of the Ly-$\alpha$ break as a redshift determinant.

Our interloper fraction is modest, and not accounting for interlopers would not have significantly affected the number densities. We note that our selection was purposely more permissive, and thus the interloper fractions that we quote are most likely higher than the true fractions of interlopers in any of the photometric luminosity function studies above, which place more emphasis on selecting a high-fidelity photometric redshift sample.

There is reasonable agreement between GOODS-North and South, although we note a difference in the median redshifts of $|\Delta z| > 0.5$ from one field to the other in both Class 4 and Class 6.1 of Medium/\jwst. This may be due to the impact of cosmic variance, and over-densities have been reported in these fields within this redshift range \citep[e.g.][]{Helton2024_overdensity}. We note that a $z\approx7.3$ over-density in GOODS-S was targeted in `GSa' which was not included in this analysis.

\section{``Gold'' samples}
\label{sec:gold_sample}

The JADES survey encompasses many different selection classes with distinct criteria, resulting in an overall inhomogeneous sample. Nonetheless, within individual tiers, the selection strategies are closely aligned. For instance, several classes across tiers select galaxies based on rest-frame UV apparent magnitude at redshifts $z>5.7$, with the main differences arising from tier-dependent depth or class-dependent redshift limits. Other classes are defined using brightness in the observed F444W filter, 
so effectively selected on the rest-optical/near-infrared luminosity. This 
provides a more reliable proxy for stellar mass than the rest-UV, which is more sensitive to ongoing star formation.\footnote{A small subset of objects are up-weighted or re-allocated as `oddballs’ and may not strictly follow these criteria. The originally assigned class is preserved in the full sample table with DR4 release (see \url{https://jades-survey.github.io/scientists/data.html}), allowing users to reverse such changes if desired.} Restricting analyses to certain priority classes across the tiers can be used to define cleaner sub-samples, as was done in the UV luminosity function comparison in Section~\ref{sec:uvlf}. However, since the original allocated classes were determined from the best photometry available at the time, and given that the underlying catalogues have since evolved (see Section~\ref{subsec:catalogues}), it is often preferable to construct homogeneous sub-samples a posteriori using updated photometry together with the measured spectroscopic redshifts. To this end, we deliver two ``gold'' samples based on spectroscopic redshifts (classes A, B, and C), applying consistent rest-UV and F444W selections using the most up-to-date photometry covering distinct redshift ranges: $z>5.7$ for the UV-selected sample and $1.5<z<5.7$ for the F444W-selected sample.  For each gold sample, we use both photometry from \citetalias{DR3}, and the most up-to-date photometry available (to be presented in Data Release 5; Robertson et al., in prep.).

The first ``gold'' sample provides rest-UV–selected galaxies over a wide redshift range. Here, photometry is measured in a 0.3\arcsec circular aperture, and for each tier we impose a magnitude cut either motivated by the faintest targets included or by the completeness limit given the 5$\sigma$ depth of the NIRCam imaging. Although these cuts were not always applied in the original targeting (which often relied on the brightest filter probing the rest-UV in \medjwst and \deepjwst), we now define a one-to-one mapping between redshift and filter, selecting the filter closest to 1500\AA\ without overlap with the Ly$\alpha$ break. The redshift-to-filter mapping is listed in Table~\ref{tab:UV_filters}, while the adopted magnitude limits for each tier are given in Table~\ref{tab:gold_UV_selection}. Although a portion of the \deepjwst footprint covers ultra-deep imaging (within the JOF), most NIRSpec slits fall on the medium-depth region of the mosaic, where the limiting depths \citep[$29.2\lesssim m_{AB}\lesssim 30$;][]{Whitler2025} lead to incompleteness in the UV-luminosity function analysis (class 6.2 point covering $29<m_{AB}<30$, see Fig.~\ref{fig:uvlf}). Accordingly, we adopt $m_{AB}<29.5$ for \medjwst and $m_{AB}<30$ for \ultradeep, while \deephst retains a brighter threshold since its selection relied primarily on \hst imaging and the early NIRCam data which did not reach the full final depths.

The second ``gold'' sample comprises galaxies selected on the basis of their F444W brightness at lower redshifts ($1.5<z<5.7$). For this sample, total fluxes are estimated using Kron apertures, and the adopted magnitude limits are given in Table~\ref{tab:gold_F444W_selection}. Since the \medhst tier was designed before NIRCam imaging became available, it cannot contribute to the F444W-based sample. By contrast, the \deephst tier had access to early \jwst imaging over much of its footprint, and the F444W-based criteria were explicitly applied to classes 7.4–7.8, allowing this tier to be included.

Tables ~\ref{tab:gold_UV_selection} and~\ref{tab:gold_F444W_selection} list the original classes within each tier that should contain these objects based on the photometry available at the time. When re-defining the ``gold'' samples using updated photometry and spectroscopic redshifts, some galaxies may be added that were not originally in these classes, while others may scatter out. For science cases that depend on the fractions of targets actually allocated to NIRSpec shutters, it is therefore important to use the spectra corresponding to the original classes, as listed in Tables~\ref{tab:priorities_Deep_JWST}--\ref{tab:priorities_Deep_HST}, since the a posteriori gold samples do not preserve the original allocation statistics.

In summary, the two ``gold'' samples provide a robust framework for analyzing the JADES dataset with homogeneous and well-defined criteria: a rest-UV–selected sample spanning a wide redshift range ($z>5.7)$, and an F444W–selected sample tracing stellar mass at lower redshifts ($1.5<z<5.7$). Together, they mitigate the complexities introduced by evolving photometry and diverse selection strategies, while preserving the flexibility to connect back to the original class-based targeting. These samples thus enable both clean statistical studies and targeted investigations of galaxy populations across cosmic time.

\begin{table}
    \centering
    \begin{tabular}{cc}
    \hline
    \hline
    Redshift Range & Filter \\
    \hline
    $z\geq9.9$ & F200W\\
    $7.2\leq z<9.9$ & F150W\\
    $5.6\leq z<7.2$ & F115W\\
    $4.6\leq z<5.6$ & F090W\\
    \hline
    \end{tabular}
    \caption{Filter chosen for UV-based gold-sample selection given the redshift (spectroscopic if available, else photometric).}
    \label{tab:UV_filters}
\end{table}

\begin{table*}
    \centering
    \caption{UV gold selection from the entire sample with $z_{spec}>5.7$ and magnitude limits based on either photometry in the \citetalias{DR3} photometric catalogues, or the most recent photometry available, labelled current.  $m_{AB}$ refers to the redshift-dependent filter choice as specified in Table~\ref{tab:UV_filters}.  The number of targets is indicated at a per-tier level, as well as the total number across the survey.  The original priority classes that are expected to contribute to these samples are indicated in the final column, though objects from other classes may be included based on final redshift and photometric measurements.}
    \begin{tabular}{ccccc}
    \hline
    \hline
    Tier & Magnitude limit & No. targets & Original classes\\
         &                 & DR3$^{a}$/current$^{b}$ \\
    \hline
    \ultradeep   & $m_{\textrm{AB}}<30$   & 20/23 & 1.1,1.2,2.1,2.3,2.4,4.1,4.2,6.1\\
    \deepjwst    & $m_{\textrm{AB}}<29.5$ & 25/27 & 1,1.1,2,3,4,6.1,6.2 \\
    \deephst     & $m_{\textrm{AB}}<29.5$ & 27/34 & 1,1.1,2,3,4,6.1,6.2 \\
    \medjwst     & $m_{\textrm{AB}}<28.5$ & 156/170 & 1,2,3,4,6.0,6.1,6.2 \\
    \medhst      & $m_{\textrm{AB}}<28.5$ & 56/81 & 1,2 \\
    \hline
    Total                   &                & 284/335 \\
    \hline
    \end{tabular}
    \label{tab:gold_UV_selection}
    
    \raggedright
    $^{a}$ Whether or not an object enters into this sample is recorded in the `UV\_gold\_DR3' column of the \href{https://jades-survey.github.io/scientists/data.html}{released catalogues}.\\
    $^{b}$ Whether or not an object enters into this sample is recorded in the `UV\_gold\_DR5\_beta' column of the \href{https://jades-survey.github.io/scientists/data.html}{released catalogues}.
\end{table*}

\begin{table*}
    \centering
    \caption{As for Table~\ref{tab:gold_UV_selection}, but now for the F444W-based ``gold'' sample.}
    \begin{tabular}{cccc}
    \hline
    \hline
    Tier & Magnitude limit & No. targets & Original classes\\
         &                 & DR3$^{a}$/current$^{b}$ \\
    \hline
    \ultradeep   & $F444W<29$   & 86/94 & 5.1,5.2,6.2,7.1,7.2,7.3,7.4\\
    \deepjwst    & $F444W<27.5$ & 66/67 & 7.5--7.9 \\
    \deephst     & $F444W<27.5$ & 75/84 & 7.5--7.8 \\
    \medjwst     & $F444W<27$ & 983/1048 & 7.5--7.9 \\
    \hline
    Total       &                & 1210/1293 \\
    \hline
    \end{tabular}
    \label{tab:gold_F444W_selection}
    
    \raggedright
    $^{a}$ Whether or not an object enters into this sample is recorded in the `F444W\_gold\_DR3' column of the \href{https://jades-survey.github.io/scientists/data.html}{released catalogues}.\\
    $^{b}$ Whether or not an object enters into this sample is recorded in the `F444W\_gold\_DR5\_beta' column of the \href{https://jades-survey.github.io/scientists/data.html}{released catalogues}.
\end{table*}

\section{Summary}
\label{sec:summary}

We have presented the target selection strategy for the JADES/NIRSpec GTO programme in the GOODS-South and GOODS-North fields.  The procedure was designed to exploit the multiplexing capability of the NIRSpec MSA, balancing rare high-redshift candidates, extreme objects, and representative galaxy populations at lower redshift. In total, more than 5,000 galaxies were assigned to shutters across the various tiers, with typical pointings yielding 150-200 spectra. 

In total, we obtain robust spectroscopic redshifts for 3,297 galaxies with a median redshift of $2.94$. 
This includes 1,231 galaxies that were selected from \emph{HST} data alone, while the remaining 2,066 were selected based on their \emph{JWST} photometry.
Across the full survey, 291 galaxies were confirmed with $z>6$, of which 50 were confirmed with $8<z_{\rm spec}\leq10$ and a further 17 were confirmed with $z_{\rm spec}>10$.


The survey spans a tiered structure of depths and modes:
\begin{itemize}
    \item \medhst: medium-depth spectra of brighter galaxies selected from \hst imaging, typically reaching m$_{\rm F160W}\lesssim$  27.
    \item \medjwst: spectra with moderately longer exposure times that were selected from NIRCam imaging.  This sample extends the dynamic range to m$_{\rm F150W} \simeq$ 28.5.
    \item \deephst and \deepjwst: extended integrations of high-redshift galaxies with 10-30 hr exposures in PRISM and shorter exposures in the gratings.
    \item \ultradeep: ultra-deep prism exposures allowing targetting of the faintest galaxies, with integrations up to $\sim50$ hrs in PRISM and $\sim37$ hrs in the gratings, providing sensitivity to continuum magnitudes of m$_{\rm F150W} \sim 30$.
\end{itemize}

We use our spectroscopic sample to construct the UV luminosity function at $z>5.7$, allowing us to account for the success rates of our photometric redshift selection.  We find this to be in good agreement with previous determinations based on purely photometric samples, suggesting that previous photometric-based determinations of the UV luminosity function have not been strongly impacted by redshift interlopers.

We also include a description of two `gold' samples that are designed to more easily explore this rich dataset with simple selection criteria.  These samples describe simple rest-UV selection at high ($z>5.7$) redshifts or rest-optical selection at lower ($1.5<z<5.7$) redshifts.

This paper is the first of two describing the final JADES/NIRSpec GTO data release, Data Release 4. Here we outline the full target selection strategy across all tiers and present the resulting redshifts and success rates. The companion paper, \citetalias{DR4_paper2}, describes the data processing and emission line measurements from the spectra.  Together, this Data Release 4 comprises the full set of NIRSpec spectra from the JADES programme, totalling 3291 galaxies with robust spectroscopic redshifts across both GOODS fields and spanning redshifts up to $z \sim 14$.  The NIRSpec component of the JADES programme demonstrates \jwst’s unique capability to push the redshift frontier, extend spectroscopic sensitivity across the near-infrared, and delivers the largest and most comprehensive survey of early galaxies to date.

\section*{Acknowledgments}

This work is based on observations made with the NASA/ESA/CSA
James Webb Space Telescope. The data were obtained from the
Mikulski Archive for Space Telescopes at the Space Telescope Sci-
ence Institute, which is operated by the Association of Universi-
ties for Research in Astronomy, Inc., under NASA contract NAS
5-03127 for JWST. These observations are associated with programmes 1210, 1180, 1181, 1286, 1287 and 3215.
The authors thank Arjen van der Wel for the comments that significantly improved the manuscript. ECL acknowledges support of an STFC Webb Fellowship (ST/W001438/1). AJB, AJC, JC \& AS acknowledge funding from the "FirstGalaxies" Advanced Grant from the European Research Council (ERC) under the European Union’s Horizon 2020 research and innovation programme (Grant agreement No. 789056).
JS, FDE, GCJ \& RM acknowledge support by the Science and Technology Facilities Council (STFC), ERC Advanced Grant 695671 "QUENCH".
FDE \& RM also acknowledge support by the Science and Technology Facilities Council (STFC), and by the UKRI Frontier Research grant RISEandFALL.
RM also acknowledges funding from a research professorship from the Royal Society.
SC \& GV acknowledges support by European Union’s HE ERC Starting Grant No. 101040227 - WINGS.
The Cosmic Dawn Center (DAWN) is funded by the Danish National Research Foundation under grant DNRF140.
CNAW, DJE, ZJ, BDJ \& MR are supported by \jwst/NIRCam contract to the University of Arizona NAS5-02015.
SA acknowledges grant PID2021-127718NB-I00 funded by the Spanish Ministry of Science and Innovation/State Agency of Research (MICIN/AEI/ 10.13039/501100011033).
DJE is also supported as a Simons Investigator. Support for program \#3215 was provided by NASA through a grant from the Space Telescope Science Institute, which is operated by the Association of Universities for Research in Astronomy, Inc., under NASA contract NAS 5-03127.
WMB gratefully acknowledges support from DARK via the DARK fellowship. This work was supported by a research grant (VIL54489) from VILLUM FONDEN.
PGP-G acknowledges support from grant PID2022-139567NB-I00 funded by Spanish Ministerio de Ciencia e Innovaci\'on MCIN/AEI/10.13039/501100011033, FEDER, UE.
BER acknowledges support from the NIRCam Science Team contract to the University of Arizona, NAS5-02105, and \jwst Program 3215. BRP acknowledges support from grant PID2024-158856NA-I00 funded by Spanish Ministerio de Ciencia e Innovaci\'on MCIN/AEI/10.13039/501100011033 and by “ERDF A way of making Europe”.
RS acknowledges support from a STFC Ernest Rutherford Fellowship (ST/S004831/1). 
ST acknowledges support by the Royal Society Research Grant G125142.
H\"U acknowledges funding by the European Union (ERC APEX, 101164796). Views and opinions expressed are however those of the authors only and do not necessarily reflect those of the European Union or the European Research Council Executive Agency. Neither the European Union nor the granting authority can be held responsible for them.
The research of CCW is supported by NOIRLab, which is managed by the Association of Universities for Research in Astronomy (AURA) under a cooperative agreement with the National Science Foundation.

\section*{Data Availability}

The datasets were derived from sources in the public domain: JWST/NIRSpec MSA and JWST/NIRCam data from MAST portal - \url{https://mast.stsci.edu/portal/Mashup/Clients/Mast/Portal.html} as well as our own reduction and analysis at \url{https://jades.herts.ac.uk/DR4/} and \url{https://jades.herts.ac.uk/search/}. All data used in this paper can be found in MAST \url{https://doi.org/10.17909/8tdj-8n28}.

\bibliographystyle{mnras}
\bibliography{bibliography.bib}

\appendix
\section{Redshift information used for prioritization}

As detailed in Sectons~\ref{subsec:classes_1_6}-\ref{subsec:class_7}, different methods were used to designate redshifts of sources for prioritization.  Table~\ref{tab:prioritisation_redshift_information} details the information available at the time and the source of the redshift information used prior to the visual inspection stage for all objects in Classes 7 and higher when selected from \jwst/NIRCam data.  It also includes \beagle and \eazy photometric redshift solutions if the quality flag of each is $<30$ along with 68\% credible intervals (see Section~\ref{subsubsec:photo_z} for further details). Objects in filler classes, or populated from HST-based catalogues are included in the full table, but with `N/A' under the `source' column.

\begin{table*}
    \caption{The redshift source used for prioritization, as well as photometric redshift information from {\scshape beagle} and \protect\textsc{eazy}.  The `source' column is either H24 for \citet{Hainline2024}, E24 for \citet{Endsley2024}, `\lbrack FILTER\rbrack\ dropout' (where \lbrack FILTER\rbrack\ gives the dropout filter for the lyman-break selection employed), `photo-z', or `spec-z'. The  \protect\textsc{eazy} and \protect\textsc{beagle} photometric redshifts are provided with 1$\sigma$ uncertainties if available, and if quality $q<30$ (see Section~\ref{subsubsec:photo_z} for details). Information of both primary and secondary peaks in the posterior probability distribution are given for \protect\textsc{beagle} photometric redshifts, though not all objects had a secondary peak. For brevity, only the first 20 objects from the top Deep/JWST priority classes are shown here, demonstrating the form and content.  The full table of over 5000 spectroscopic targets will be available in machine-readable form once published.}
    \centering
    \begin{tabular}{cccccccccc}

\hline
     &             &          &        &                & \beagle         & \beagle           \\
Tier & NIRSpec\_ID & Priority & Source &  \eazy photo-z & primary photo-z & secondary photo-z \\
\hline
\deepjwstgs & 183348 & 1 & H24 & 14.4$^{+0.6}_{-0.6}$ & 15.0$^{+0.9}_{-0.9}$ & -\\
\deepjwstgs & 183349 & 1 & nearby companion & 3.7$^{+0.2}_{-0.3}$ & 3.4$^{+0.2}_{-0.2}$ & -\\
\deepjwstgs & 20012702 & 1 & H24 & 10.9$^{+0.6}_{-0.3}$ & - & - \\
\deepjwstgs & 20013731 & 1 & H24 & 14.0$^{+0.5}_{-1.0}$ & - & - \\
\deepjwstgs & 20018044 & 1 & H24 & 14.4$^{+0.8}_{-1.4}$ & - & - \\
\deepjwstgs & 20055733 & 1 & H24 & 14.4$^{+1.0}_{-1.2}$ & - & - \\
\deepjwstgs & 20064312 & 1 & H24 & 10.6$^{+0.7}_{-0.4}$ & - & - \\
\deepjwstgs & 20176151 & 1 & H24 & 11.0$^{+0.3}_{-0.3}$ & - & - \\
\deepjwstgs & 20015720 & 1.1 & H24 & 11.5$^{+0.6}_{-0.5}$ & - & - \\
\deepjwstgs & 20177294 & 1.1 & H24 & 10.5$^{+0.03}_{-1.0}$ & - & - \\
\deepjwstgs & 9442 & 2 & F090W dropout & 1.7$^{+5.1}_{-0.2}$ & 1.5$^{+0.1}_{-0.1}$ & 6.9$^{+0.8}_{-0.8}$\\
\deepjwstgs & 12326 & 2 & photo-z & 8.3$^{+0.3}_{-0.4}$ & 8.3$^{+0.7}_{-0.7}$ & - \\
\deepjwstgs & 17909 & 2 & F115W dropout & 4.2$^{+4.1}_{-2.7}$ & 2.1$^{+0.6}_{-0.6}$ & 9.2$^{+0.8}_{-0.8}$\\
\deepjwstgs & 20005936 & 2 & H24 & 8.0$^{+0.2}_{-0.2}$ & - & - \\
\deepjwstgs & 20006347 & 2 & H24 & 8.8$^{+0.1}_{-0.1}$ & - & - \\
\deepjwstgs & 20015285 & 2 & H24 & 9.4$^{+0.1}_{-7.0}$ & - & - \\
\deepjwstgs & 20021387 & 2 & H24 & 10.9$^{+0.8}_{-0.6}$ & - & - \\
\deepjwstgs & 20050575 & 2 & H24 & 8.5$^{+0.1}_{-0.1}$ & - & - \\
\deepjwstgs & 20051718 & 2 & H24 & 8.0$^{+0.2}_{-0.1}$ & - & - \\
\deepjwstgs & 20062446 & 2 & H24 & 8.3$^{+0.3}_{-0.2}$ & - & - \\
\hline
\label{tab:prioritisation_redshift_information}
\end{tabular}
\end{table*}

\label{lastpage}

\end{document}